\documentclass[]{JHEP}

\def\bea{\begin{eqnarray}}
\def\eea{\end{eqnarray}}
\def\be{\begin{equation}}
\def\ee{\end{equation}}
\def\nn{\nonumber}

\def\Z{{\bf Z}}

\title{Sigma-model symmetry in orientifold models}

\author{Claudio A. Scrucca \\
Sektion Physik, Ludwig Maximilian Universit\"at M\"unchen\\
Theresienstrasse 37, 80333 Munich, Germany\\
{\footnotesize \tt Claudio.Scrucca@physik.uni-muenchen.de}}
\author{Marco Serone \\
Department of Mathematics, University of Amsterdam \\
Plantage Muidergracht 24, 1018 TV Amsterdam, The Netherlands; \\
Spinoza Institute, University of Utrecht \\
Leuvenlaan 4, 3584 CE Utrecht, The Netherlands \\
{\footnotesize \tt serone@wins.uva.nl}}

\abstract{ 
We investigate in the simplest compact $D=4$ $N=1$ Type IIB orientifold 
models the sigma-model symmetry suggested by the proposed duality of 
these models to heterotic orbifold vacua. This symmetry is known
to be present at the classical level, and is associated to a composite
connection involving untwisted moduli in the low-energy supergravity theory.
In order to study possible anomalies arising at the quantum level, we 
compute potentially anomalous one-loop amplitudes involving gluons, 
gravitons and composite connections. We argue that the effective vertex 
operator associated to the composite connection has the same form as that 
for a geometric deformation of the orbifold. Assuming this, we are able
to compute the complete anomaly polynomial, 
and find that all the anomalies are canceled through a Green-Schwarz 
mechanism mediated by twisted RR axions, as previously conjectured.
Some questions about the field theory interpretation of our results
remain open.}


\preprint{UvA-WINS-Wisk-00-09 \\ SPIN-2000/18 \\ LMU-TPW 00-16  \\ 
{\tt hep-th/0006201}}

\begin{document}


\section{Introduction}

Recently, renewed interest has been devoted to orientifold 
vacua of Type IIB string theory, constructed by projecting 
out a standard toroidal compactification by the combined 
action of a discrete spacetime orbifold symmetry $G$ and the 
world-sheet parity $\Omega$ \cite{sag,ps,hor,dlp}. These 
unoriented string theories contain both open and closed 
strings, and constitute the perhaps most important and concrete 
example of models in which gauge interactions are localized on 
D-branes \cite{pol}. They are therefore the natural arena for 
the realization of the ``brane-world'' scenario. Furthermore, 
this kind of models have proven to offer surprisingly 
attractive possibilities from a phenomenological point of 
view (see for instance \cite{aiq,aiqu}). 

In the following, we will be concerned with compact $D=4$ $N=1$ 
Type IIB orientifold models \cite{abpss,ks,afiv,z,kst}. 
These vacua represent a simple and
tractable prototype of more general and possibly non-supersymmetric 
orientifold models. Some of them are also phenomenologically appealing
and constitute a viable alternative to their more 
traditional heterotic analogues. 
In fact, a weak-weak Type IIB - heterotic duality has been conjectured 
\cite{abpss,afiv,kst,afiuv} for several pairs of 
vacua\footnote{Notice that, although 
Type IIB $\!/\Omega$ = Type I \cite{sag}, this duality is not a trivial
consequence of Type I - heterotic duality in $D=10$ \cite{polwit}, 
because in general $\Omega$ and $G$ do not commute (see \cite{pol2} for 
a discussion in the $D=6$ models of \cite{bs,gp,gj,dp}), 
and therefore Type IIB $\!/\{\Omega,G\} \neq$ Type I $\!/G$.}. In particular, 
$\Z_N$ orientifolds with $N$ odd do contain D9-branes but no D5-branes,
and could be dual to the corresponding perturbative $\Z_N$ heterotic
orbifold. Models with $N$ even do instead contain both D9-branes and 
D5-branes, and could be dual to heterotic orbifolds with a perturbative 
sector corresponding to D9-branes and a non-perturbative instantonic 
sector corresponding to D5-branes \cite{afiuv}. 

At the classical level, evidence for the duality is suggested by the 
almost perfect matching of the low-energy spectra and the fact
that the orientifold models seem to possess the same classical symmetries 
as their heterotic companions \cite{iru2}. In particular, they both 
possess a so-called ``sigma-model'' symmetry\footnote{In the following, we 
shall often use the abbreviation ``sigma'' for ``sigma-model''.}, 
naturally emerging from 
$N=1$ supergravity. More precisely, this symmetry consists of $SL(2,R)$ 
transformations for the untwisted $T^i$ moduli and the other chiral 
superfields, implemented as the combination of a K\"ahler transformation
and a reparametrization of the scalar K\"ahler manifold.
On the heterotic side, a discrete $SL(2,Z)$ subgroup of these transformations
is known to correspond to the well-known T-duality symmetry, valid to all 
orders of string perturbation theory, and is therefore expected to be exact. 
On the orientifold side, instead, sigma-model transformations do not seem 
to correspond to any known underlying string symmetry, and it is not clear 
whether the symmetry is exact. At the quantum level, the comparison 
becomes much more involved and several subtleties arise. 
In particular, it has been argued in \cite{abd} that the one-loop 
corrected gauge couplings in orientifold models seem to be incompatible 
with any duality map (see also \cite{lln} for further discussion).
There is however an important issue which can be addressed even without 
knowing the detailed duality map: whether or not the classical sigma-model
symmetry is anomalous at the quantum level. 
The latter continuous symmetry is indeed associated to a composite connection
in the low-energy effective supergravity theory, and acts as chiral rotations
on all the fermions. There are therefore anomalous triangular diagrams 
involving gluons, gravitons and composite connections, leading in general
to a non-vanishing one-loop anomaly. On the heterotic side, this one-loop 
anomaly is canceled by a universal tree-level Green-Schwarz (GS) mechanism 
mediated by the dilaton \cite{bg,dfkz}, and the appearance of the appropriate
GS term has been explicitly checked through a string theory computation
\cite{agnt}. On the orientifold side, it was proposed in \cite{iru2} that 
a similar GS mechanism involving both the dilaton and
twisted RR axions could cancel the anomalies. 
This observation was motivated by the factorizability of mixed 
sigma-gauge anomalies computed from the low-energy spectra. 
However, it was subsequently argued in \cite{lln} that requiring a 
similar mechanism also for mixed sigma-gravitational anomalies 
would lead to an apparent contradiction with the known results for 
gauge-gravitational anomaly cancellation \cite{ss3}. 
The question of whether the sigma-model symmetry is 
anomalous or not in orientifold models is therefore still unclear
and of extreme relevance for their duality to heterotic theories.
Note however that even if the presence of anomalies would pose serious
problems to the duality, it would not be fatal 
for the consistency of the 
orientifold models in themselves\footnote{Even in the worse case in which 
all types of mixed anomalies arise, it is always possible to redefine
the conserved currents and energy-momentum tensor in such a way to
eliminate mixed gauge or gravitational anomalies, and push all the anomaly 
in the sigma-model symmetry only. The latter is not fatal, since the emerging 
longitudinal states are composite and not elementary, and so cannot
violate unitarity in higher-loop diagrams as would do longitudinal 
gluons or gravitons resulting from gauge or gravitational anomalies.}.

The aim of this paper is to study the cancellation of all possible 
(pure or mixed) sigma-gauge-gravitational anomalies in orientifold
models through a string theory computation. Such an 
analysis is interesting by itself even beyond the context of 
Type IIB - heterotic duality, since it can provide useful informations 
about the low-energy effective action. For instance, the GS couplings 
that will be derived are related by supersymmetry to other couplings 
in the Lagrangian and determine under suitable assumptions the gauge 
kinetic functions and the Fayet-Iliopoulos terms. For simplicity, 
the analysis will be restricted to the models with $N$ odd. These 
are indeed simpler than models with $N$ even for a variety of reasons;
in particular, they do not present threshold corrections \cite{abd}.
The only consistent models with $N$ odd are the $\Z_3$ and the $\Z_7$ 
models.

We follow the strategy developed
in \cite{ss2,ss3} for standard gauge-gravitational anomalies, and 
compute both the quantum anomaly and the classical inflow in all
possible channels. By factorization, it is then possible to extract
all the anomalous couplings for D-branes and fixed-points present
in each model, and the GS term given by their sum.
A major ingredient of our computation is an effective vertex operator
for the composite sigma-model connection, which results from a pair 
of untwisted K\"ahler moduli. We provide arguments that such a vertex 
is in fact the same as that of an ``internal graviton'' associated 
to a deformation of the K\"ahler structure of the orbifold respecting 
its rigid complex-structure. This suggests that there is a close relation 
between sigma-model symmetry and invariance under reparametrizations of 
the internal part of the spacetime manifold. In particular, potential 
anomalies in these symmetries seem to coincide. 
Assuming the relation above to be valid and
using this common vertex, we are able to compute the complete anomaly 
polynomial as a function of the gauge, gravitational and sigma-model 
curvatures. We find that all the anomalies are canceled through 
a GS mechanism mediated by twisted RR axions only, extending the 
results of \cite{ss3} for gauge-gravitational anomalies. 
The dilaton does not play any role in the anomaly cancellation mechanism,
contrarily to what proposed in \cite{iru2} and in agreement with
\cite{lln}.

The results of our string computation disagree with the field theory
analysis of \cite{lln} on a crucial 
sign in the contribution of the twisted modulini to the one-loop 
sigma-gravitational anomaly. 
Contrary to what assumed in \cite{lln}, 
it seems that these twisted closed string states must have a 
non-vanishing ``effective'' modular weight, that is responsible
for the full cancellation of all the anomalies. Although we do not 
have yet a complete understanding of the field theory interpretation 
of our results and their implications on the low-energy effective action, 
we believe that they rise some questions about the actual form of 
the K\"ahler potential for twisted fields. 
As far as we know, this potential has not yet been unambiguously determined. 
The only available proposal about its form is that of \cite{popp}, 
and it was indeed assumed in \cite{lln}. However, this potential 
implies vanishing modular weight for twisted fields, in apparent 
contradiction with our results, at least if one does not include 
possible tree-level corrections to it induced by the GS mechanism.
Whether our string results might be explained by taking into account
the GS terms in the potential proposed in \cite{popp},
or they imply a different form for the K\"ahler potential
of twisted fields, has still to be understood \cite{kiss}.
 
Independently of their actual field theory explanation, we think 
that our results provide strong evidence for the occurrence of this
cancellation mechanism, generalizing it moreover to all the other 
types of anomalies, like in particular pure sigma-model anomalies. 
Unfortunately, although we provide several convincing arguments 
on the correctness of the effective vertex operator for
the composite sigma-model connection, a rigorous proof is missing.
Therefore, the only safe statement that we are in position to 
make is that the associated symmetry is preserved at the 
one-loop level thanks to a generalized GS mechanism. Whether or 
not this is really the sigma-model symmetry remains strictly speaking
to be proven, although we believe that it is quite unlikely that 
this is not the case since all the anomalies we compute have 
precisely the structure expected for sigma-gauge-gravitational 
anomalies. Notice also that thanks to the alternative interpretation 
of this vertex as an internal graviton, these canceled anomalies 
can be inequivocably interpreted as relative to internal reparametrizations. 
As such, they admit a topological interpretation in terms of equivariant 
indices of the spin and signature complexes, and it is possible to verify 
the results obtained through the direct string computation by applying 
suitable index theorems, as we will see. 

The structure of the paper is the following. In Section 2, we briefly 
review the notion of sigma-model symmetry. In Section 3, we set up the 
general strategy of the string computation and propose a possible 
path-integral derivation of the effective vertex for 
the composite connection. In Section 4 we perform the string computation
on the four surfaces appearing at the one-loop order. In Section 5, we
reproduce the same results from a mathematical point of view as topological
indices. In Section 6, we discuss in more detail the obtained quantum 
anomalies and perform the factorization of the classical inflow to get
all the RR anomalous couplings and the total GS term. In Section 7, we 
discuss possible field-theory interpretations of our results and their
implications. Finally, we give conclusions in Section 8. In Appendix A,
we report useful conventions about $\vartheta$-functions. In Appendix B,
we discuss the cancellation of anomalies in Type IIB string theory
(this completes the analysis in \cite{ss2}). Finally, Appendix C
contains some useful details about the string computation.


\section{Sigma-model symmetry}

In this section, we briefly review some well-known facts about the 
sigma-model symmetry, and discuss its potential anomalies in $D=4$, $N=1$ 
supergravity models. These general concepts are useful for the considerations 
that will follow, in particular in Section 7.

The scalar manifold ${\cal M}$ of any generic $D=4$ $N=1$ supergravity model 
is known to be a K\"ahler manifold, described by a K\"ahler potential
$K$. At the classical level the Lagrangian presents two distinct symmetries
(beside possible local gauge symmetries):
\begin{itemize}
\item{K\"ahler symmetry, under which the K\"ahler potential transforms 
as\footnote{The Lagrangian is invariant under (\ref{K}) if also the
superpotential $W$ transforms as $W\rightarrow e^{-F} W$. Since $W$
is irrelevant in the considerations that will follow, we will neglect it.}
\be
\kappa^2 K(\Phi^M,\bar\Phi^M) \rightarrow 
\kappa^2 K(\Phi^M,\bar\Phi^M) + F(\Phi^M)
+\bar F(\bar\Phi^M) \;. \label{K}
\ee}
\item{Global isometries of ${\cal M}$, under which
\be
\phi^M \rightarrow \phi^{\prime M}(\phi^N) \;. \label{sigma}
\ee}
\end{itemize}
Here $\Phi^M$ and $\phi^M$ denote all the chiral multiplets
in the model and their lowest components, $F(\Phi)$ is a generic
chiral superfield, and $\kappa^2$ is Newton's gravitational constant.  
The fermions $\psi^M$ in the chiral multiplet $\Phi^M$
transform also under (\ref{K}) and (\ref{sigma}). 
Correspondingly, the fermionic kinetic terms contain a 
covariant derivative involving the following ``K\"ahler'' and
``isometry'' connections \cite{cfgp}:
\bea
&& A_\mu^{(K)}=-\frac i2 \kappa^2 \, ( \, K_M \, \partial_\mu \phi^M
- K_{\bar M} \, \partial_\mu \phi^{\bar M} \, ) \;, \label{Kconn} \\
&& A_{\mu\;\;\;\;\; N}^{(I)\, M} = i \, ( \Gamma^M_{\, KN} \, \partial_\mu
\phi^K - \Gamma^{\bar M}_{\, \bar K \bar N} \, \partial_\mu \phi^{\bar K})
\;. \label{Sconn}
\eea
Here $\phi^{\bar M} \equiv \bar \phi^M$, $K_M$ and $K_{\bar M}$ denote
the derivative of $K$ with respect to the corresponding fields and 
$\Gamma$ is the usual Christoffel connection on the K\"ahler manifold 
${\cal M}$. Notice that the above connections are not new fundamental 
states, but composites of the scalar fields.

At the quantum level, the symmetries associated to (\ref{K}) and 
(\ref{sigma}) might be spoiled by triangular one-loop graphs involving 
as external states the connections (\ref{Kconn}) and (\ref{Sconn}), as 
well as gluons or gravitons. A direct evaluation of these mixed 
anomalies is not an easy issue, because of the compositeness of the 
connections. One can however use indirect arguments 
that rely on the similarity of the structure of the associated anomalous 
one-loop amplitudes with that of standard $U(1)$-gauge and
$U(1)$-gravitational anomalies \cite{bg,dfkz}. 
We shall briefly review this analogy in the following, 
focusing on the case in which a single composite connection enters as 
external state in the anomalous diagram.

The considerations made so far are quite general and apply to any $D=4$ 
$N=1$ model. We specialize now to the low-energy Lagrangians arising 
from the Type IIB orientifolds we want to analyze, {\it i.e.} the $\Z_3$ 
and the $\Z_7$ model (see \cite{abpss,ks,afiv} for more details on 
these string vacua). The massless closed string spectrum of these 
models contain the gravitational multiplet, a universal chiral multiplet 
$S$, three chiral multiplets $T^i$ corresponding to the (complexified) 
K\"ahler deformations of the three internal two-tori\footnote{Actually,
additional ``off-diagonal'' untwisted moduli survive the orientifold 
projection in the special $\Z_3$ model. We do not consider them here 
for simplicity, and all the considerations that follow are 
independent of the presence of these fields.}, 
and a given number of chiral multiplets $M^{\alpha}$ 
arising from the twisted sectors of the orbifold. The open string 
spectrum (from D9 branes only in these models) contains vector 
multiplets and three groups of charged chiral multiplets $C^a$. 
In order to distinguish the different coordinates of ${\cal M}$, we use 
the index $M=(i,a,\alpha)$ for $T^i,C^a$ and $M^\alpha$ respectively.
As we will see in next sections, the dilaton field $S$ does not
participate at all to the GS mechanism canceling the anomalies,
and is inert under any gauge, diffeomorphism or sigma-model
transformations.

Up to quadratic order in the charged fields, the total K\"ahler potential 
of these orientifolds is believed to be \cite{iru2}\footnote{See also 
\cite{bb} for further considerations on the K\"ahler potential
of $D=4$ orientifold models.}:
\bea
\kappa^2 \, K_{tot}(\Phi^M,\bar\Phi^M) &=& - \ln (S+\bar S) -
\sum_{i=1}^3 \ln (T^i+\bar T^i) +  \sum_{i=1}^3 \delta^a_i 
\frac {\bar C^a C^a}{T^i+\bar T^i}  \nn \\
&\;& + \, \kappa^2 \, K^{(M)}(M^\alpha,\bar M^\alpha,T^i,\bar T^i) \;,
\label{Ktot}
\eea
where $K^{(M)}$ is an unknown potential for the twisted fields
$M^\alpha$. As mentioned in the introduction, the sigma-model 
symmetry we want to study in these orientifold models is 
the dual of heterotic T-duality. It acts on the fields $T^i$, 
$C^a$ and $M^\alpha$ through the following $SL(2,R)_i$ 
transformations (no sum over $i$, $ad-bc=1$):
\bea
&& T^i \rightarrow \frac {a_i T^i - i b_i}{i c_i T^i + d_i} 
\;, \label{sl2r} \\
&& C^a \rightarrow \frac {\delta^a_i}{(ic_iT^i + d_i)} \, C^a 
\;, \label{Ctr} \\
&& M^\alpha \rightarrow M^{\prime \alpha}(M^\beta,T^i) \label{Mtr} \;,
\raisebox{16pt}{}
\eea
and similarly for the complex conjugate fields. 
The transformation (\ref{Ctr}) leaves the corresponding (third) term 
of the K\"ahler potential (\ref{Ktot}) invariant and
(\ref{Mtr}) is chosen in such way to preserve the 
last contribution $K^{(M)}$. On the other hand, (\ref{sl2r}) produces 
a non-trivial transformation of the second term. In total, 
the complete K\"ahler potential (\ref{Ktot}) undergoes the following 
K\"ahler transformation under (\ref{sl2r}), (\ref{Ctr}) and (\ref{Mtr}):
\be
\kappa^2 \, K_{tot}(\Phi^M,\bar\Phi^M) \rightarrow \kappa^2 \, 
K_{tot}(\Phi^M,\bar\Phi^M) + \lambda^i(T^i) + \bar \lambda^i(\bar T^i) \;,
\label{Ktr}
\ee
with
\be
\lambda^i(T^i) = \ln (ic_iT^i+d_i) \;.
\label{lambda}
\ee
The sigma-model symmetry in question is therefore the combination of 
an isometry and a K\"ahler transformation, and potential anomalies
will therefore involve both connections (\ref{Kconn}) and (\ref{Sconn}).

In order to be able to derive an explicit formula at least for 
mixed sigma-gauge/gravitational anomalies, we need to make some 
extra assumptions on the potential $K^{(M)}$ and the transformations 
(\ref{Mtr}). We take here the one usually considered in the literature, 
that indeed holds generically for heterotic models \cite{dkl}:
\bea 
&& \kappa^2 K^{(M)}(M^\alpha,\bar M^\alpha,T^i,\bar T^i) =
\sum_{\alpha} \prod_{i=1}^3 (T^i+\bar T^i)^{n^\alpha_i} 
\bar M^\alpha M^\alpha + ...  \;, \nn \\
&& M^\alpha \rightarrow (ic_iT^i+d_i)^{n^\alpha_i} M^\alpha  
\;,  \label{modw}
\eea
where the dots stand for possible higher order terms in 
$M^\alpha,\bar M^\alpha$. The numbers $n^\alpha_i$ are the 
so-called ``modular weights'' \cite{il} of the fields $M^\alpha$.
It is straightforward to see that for the reparametrizations
(\ref{sl2r}), (\ref{Ctr}) and (\ref{Mtr}), and the K\"ahler 
transformation (\ref{Ktr}) and (\ref{lambda}) ($F = \lambda^i$), 
the total connection 
$Z_\mu^M \equiv A_\mu ^{(K)} + A_{\mu\;\;\;\;\;M}^{(I)\, M}$
transforms as a $U(1)$ connection\footnote{In deriving
(\ref{cos}) we assumed that the orbifold limit corresponds to 
$\langle C^a \rangle = \langle M^\alpha \rangle = 0$.
The orbifold limit, however, is generically assumed to be given by
$\langle m^\alpha \rangle =0$, where the scalars $m^\alpha$ belong
to the linear multiplets $L^\alpha$, dual of the chiral multiplets
$M^\alpha$ \cite{abd}. So we are assuming that at leading order
$\langle m^\alpha \rangle = 0$ corresponds to 
$\langle M^\alpha \rangle =0$.}:
\be
Z_\mu^M \rightarrow Z_\mu^M \, + \, (1+2n_i^M) \, 
\partial_\mu \, {\rm Im} \lambda^i \;,
\label{cos}
\ee
where $n_i^\alpha$ are the coefficients defined in (\ref{modw}), 
$n^a_i = - \delta^a_i$, and $n^j_i=-2\, \delta^j_i$. 
The sigma-model symmetry can therefore be viewed as a $U(1)_i$ symmetry 
with ``modular charge'' $Q_i^M = (1+2n_i^M)$.
The explicit form of $Z^M$ and its field-strength $G^M=dZ^M$ can be
easily evaluated. It is actually convenient to disentangle the modular
charges $Q_i^M$ from the connection and define the three connections
$Z_{\mu,i}$ and their field-strength $G_{\mu,i}$ so that
$Z_\mu^M = \sum_i Q_i^M Z_{i,\mu}$ and 
$G_{\mu\nu}^M = \sum_i Q_i^M G_{i,\mu\nu}$.
One finds:
\bea
&& Z_{i,\mu} = \frac i2 \,\frac 
{\partial_\mu (t^i - \bar t^i)}{t^i+\bar t^i} \;, \label{FKexc} \\
&& G_{i,\mu\nu} = 2i \,
\frac {\partial_{[\mu} t^i \partial_{\nu ]}\bar t^i}
{(t^i+\bar t^i)^2} \;. \label{FKex}
\eea
Sigma-gauge/gravitational anomalies can then be computed 
by treating them as $U(1)_i$-gauge/gravitational anomalies 
(in the following denoted briefly by $FFG_i$ and $RRG_i$ 
anomalies respectively). 
Explicit formulae for the anomaly coefficients can be found 
for example in eqs.(2.8) and (2.12) of \cite{iru2}.


\section{Anomalies in orientifold models}

In this section, we will set up the general strategy for studying
all types of anomalies in chiral orientifold models, and 
investigate their cancellation. We will begin by reviewing the 
main aspects of the approach developed in \cite{ss2,ss3} for 
standard anomalies (see also \cite{bm} for a similar analysis in 
non-geometric models), and generalize it to sigma-gauge-gravitational 
anomalies.

To begin, we shall briefly recall some basic but important  
facts about anomalies for the convenience of the reader. Anomalies 
in a quantum field theory effective action have to satisfy the 
Wess-Zumino (WZ) consistency condition. These in turn imply that 
any anomaly in $D$ dimensions is uniquely characterized by a 
gauge-invariant and closed $(D+2)$-form $I$. 
Using the standard WZ-descent notation\footnote{The invariant 
closed $(D+2)$-form $I$ defines locally a non-invariant Chern-Simons 
$(D+1)$-form $I^{(0)}$ such that $I = d I^{(0)}$, whose gauge variation 
then defines a $(D)$-form $I^{(1)}$ through $\delta I^{(0)} = d I^{(1)}$.}: 
${\cal A} = 2\pi i \int I^{(1)}$.
The anomaly polynomial $I$ is a characteristic class of the gauge 
and tangent bundles, of degree $(D+2)/2$ in the curvature two-forms.

\subsection{The strategy}

The cancellation of anomalies in string theory is achieved in a very 
natural and elegant way, and is intimately related to more general 
consistency requirements, like modular invariance 
and tadpole cancellation. Possible anomalies arise exclusively from 
boundaries of the moduli space of one-loop string world-sheets.
Moreover, direct computations have shown \cite{GS} that the whole 
tower of massive string states contribute in general to anomalies 
in such a way that these vanish for consistent models, even if 
the massless spectrum is generically anomalous on its own.
From a low-energy effective field theory point of view, where
massive states are integrated out and only the resulting effective 
dynamics of the light modes is considered, the total one-loop anomaly 
is canceled by an exactly opposite anomaly arising in tree-level 
processes involving the magnetic exchange of tensor fields \cite{GS2}. 
This is the celebrated Green-Schwarz (GS) mechanism \cite{GS2}, and is
an absolutely crucial ingredient for the existence of consistent
supersymmetric chiral gauge theories in higher dimensions.

In the following, we will focus on the $CP$-odd part of the one-loop 
effective action, where anomalies arise. For consistent models, the 
exact string theory computation is expected to yield a vanishing anomaly. 
However, as discussed above, this is interpreted as a non-trivial GS mechanism 
of anomaly cancellation in a low-energy effective theory valid
at energies $E \ll 1/\sqrt{\alpha^\prime}$. In order to get 
directly this low-energy approximation, one can take the limit 
$\alpha^\prime \rightarrow 0$ from the beginning, before 
integrating over the world-sheet moduli. 
The motivation to pursue this strategy, 
instead of the more direct full string theory computation, is threefold. 
First, the required computations simplify dramatically. Furthermore, 
one gets an improved understanding of the low-energy mechanism of anomaly 
cancellation. Finally, one can extract important WZ
couplings appearing in the effective action by factorization \cite{ss2,ss3}.

Consider now orientifold models. The relevant anomalous string diagrams 
are the annulus ($A$), the M\"obius strip ($M$) and the Klein bottle ($K$). 
These world-sheet surfaces lead to potential divergences due to possible
tadpoles for massless particles propagating in the transverse channel.
Consequently, they also lead to potential anomalies.
In addition, also the torus ($T$) surface can be anomalous, in the limit 
under consideration. 
We will see that there are contributions to the anomaly from this diagram, 
but they turn out to always cancel among themselves. 

The most general situation which is allowed by the property that
anomalous amplitudes are boundary terms in moduli space is the following. 
The $A$, $M$ and $K$ surfaces are parametrized by a real modulus 
$t \in [0,\infty]$. The contribution from the boundary at 
$t \rightarrow \infty$ is interpreted as the standard quantum anomaly, 
whereas the contributions from the other boundary at $t \rightarrow 0$ 
is interpreted as classical inflow of anomaly. The $T$ amplitude 
is instead parametrized by a complex modulus $\tau \in {\cal F}$, where 
${\cal F}$ is the fundamental domain. Again, the contribution from the 
component $\partial {\cal F}_{\infty} = [-1/2 + i\,\infty, 1/2 + i\,\infty]$ 
of the boundary $\partial {\cal F}$ at infinity is interpreted as the 
standard quantum anomaly, whereas the contribution from the remaining 
component $\partial {\cal F}_0$ should be associated to the classical 
inflow of anomaly. Summing up, one would therefore get a quantum anomaly 
${\cal A} = (A+M+K+T)|_\infty$ and a classical inflow 
${\cal I} = (A+M+K+T)|_0$. It should be however mentioned that the 
above interpretation for the $T$ surface involves some conceptual 
subtleties related to modular invariance, that might mix different 
contributions.
Luckily, we will see that the $T$ amplitude gives a vanishing 
contribution anyhow: the pieces in the $\partial {\cal F}_0$ component 
cancel pairwise thanks to modular invariance \cite{susu}, that still
holds in the $\alpha^\prime\rightarrow 0$ limit, 
whereas the $\partial {\cal F}_\infty$ 
component vanishes by itself. Moreover, the $A$, $M$ and $K$ 
contributions are topological and independent of the modulus. 
Correspondingly, ${\cal A}$ and ${\cal I}$ are identical to each 
other and cancel.

As last important remark, notice that in four dimensions even
in non-planar diagrams the closed string state exchanged
in the transverse channel is always on-shell, due to the 
conservation of momentum. Strictly speacking, this means that 
the usual argument for the cancellation of anomalies at 
the string level \cite{GS} does not apply in this case,
giving further motivation for a detailed analysis.

\subsection{Set-up of the computation}

The computation of the $A$, $M$, $K$ and $T$ amplitudes proceeds along
the lines of \cite{ss2,ss3}, that we shall briefly review and extend.
For the time being, we shall assume that the composite connections
(\ref{FKexc}) are described by suitable effective vertex operators, 
postponing a detailed discussion of this issue to next subsection.

An anomaly of the type discussed above, in the $CP$-odd part of the
effective action, is encoded in a one-loop correlation function in the 
odd spin-structure on the $A$, $M$ and $K$ surfaces, and in the 
odd-even and even-odd spin-structures on the $T$ surface, involving 
gluons, gravitons and composite connections. Denoting by 
$\rho$ the modulus of the surface and by ${\cal F}$ its integration 
domain, one has on a given surface and spin-structure
\be
{\cal A}_{1...n} = \int_{\cal F} \! d\rho \, 
\langle {V_1}^\prime V_2... V_{n} \, J \rangle \;.
\ee
The insertion of the supercurrent $J$ is due to the existence of a 
world-sheet gravitino zero-mode; more precisely, $J = T_F + \tilde T_F$ 
in the odd spin-structure on the $A$, $M$ and $K$ surfaces, and 
$J = T_F, \tilde T_F$ in the odd-even and even-odd spin-structures 
respectively on $T$. The vertex $V^\prime$ is taken in the $-1$-picture in 
the odd sector and represents an unphysical particle. Taking the 
latter to be a longitudinally polarized gluon, graviton or composite 
connection, one computes the variation of the one-loop effective
action under gauge, diffeomorphisms or sigma-model transformations. 
The remaining vertices $V$ are taken in the $0$-picture and represent 
physical background gluons, gravitons or composite connections. 
Thanks to world-sheet supersymmetry and the limit 
$\alpha^\prime \rightarrow 0$, one can use effective vertex operators
which are simpler to handle. 

After some formal manipulations, the correlation function above 
can be rewritten as boundary terms in moduli space \cite{ikk,susu} 
\be
{\cal A}_{1...n} = \oint_{\partial \cal F} \! d\rho \,
\langle W_1 V_2 ... V_n \rangle \; ,
\label{Ceff}
\ee
where $W$ is an auxiliary vertex defined out of $V^\prime$ for the 
unphysical particle. Importantly, the vertices $V$'s contain two tangent 
fermionic zero-modes, whereas $W$ does not contain any of them.
The insertion of $W$, rather than $V$, for the unphysical particle 
representing the gauge variation of the one-loop effective action 
corresponds to the fact that the anomaly ${\cal A}$ is given by the 
WZ descent of the anomaly polynomial $I$: $A = 2 \pi i \int I^{(1)}$. 
More precisely, one can show \cite{ss2,ss3} that the latter is obtained
simply by substituting back $V$ instead of $W$, that is 
\be
I_{1...n} = \oint_{\partial \cal F} \! d\rho \,
\langle V_1 V_2 ... V_n \rangle \;,
\ee
with the convention of working in two more dimensions and omitting the 
integration over bosonic zero-modes.
Finally, it is possible to define the generating functional of all
the possible anomalies by exponentiating one representative vertex 
for each type of particle and compute the resulting deformed partition
function $Z'$.
Finally, the total anomaly polynomial is given just by
\be
I = \oint_{\partial \cal F} \! d\rho \, Z^\prime \;.
\label{I}
\ee

\subsection{Effective vertices}

The fact that one can use effective vertices in the computation of 
the partition function yielding the anomaly polynomial is due to the
$\alpha^\prime \rightarrow 0$ limit and to certain special properties 
of correlation functions in supersymmetric spin-structures like those 
of relevance here. One way to understand this is to notice that the 
partition functions to be computed are related to topological indices 
which are almost insensitive to any continuous parameter deformation. 
From a more technical point of view, there is 
always a fermionic zero-mode for each spacetime direction. 
The corresponding Berezin integral in the partition function yields
a vanishing result unless the interaction vertices provide one of 
each fermionic zero-mode. Infact, products of these fermionic zero 
modes provide a basis of forms of all degrees in the target spacetime,
the Berezin integral selecting the appropriate total degree.

On general grounds, it is expected that the effective vertices
depend only on the corresponding curvature. Since these behave 
as two-forms, they must be contracted with two tangent fermionic 
zero-modes. Moreover, the vertices must be world-sheet supersymmetric. 
Finally, thanks to the $\alpha^\prime \rightarrow 0$ limit, they cannot 
contain additional momenta, beside from those defining the curvature.
These three basic requirements, together with the index structure 
of the curvatures and conformal invariance, turn out to severely 
constrain the effective vertices in each case. 
For gluons and gravitons, they can be derived in a 
straightforward way as in \cite{ss2}, but for the composite connections
(\ref{FKexc}), the analysis is much more involved since the latter are not 
fundamental fields but composite of the scalar fields of the theory, 
and there are therefore no vertex operators directly associated to them. 
Our main observation is that the field-strengths (\ref{FKex}) have a 
quadratic dependence on the untwisted $t^i$ and $\bar t^i$ moduli
fluctuations. Correspondingly, suitable amplitudes with the insertion 
of the vertex operators associated to these scalars should reproduce
the insertion of the composite connections (\ref{FKexc}).
The untwisted $t^i$ moduli are defined as \cite{afiv}
\be
t^i = e^{-\phi_{10}}\, g_{i\bar i} \, + \, i \theta_i \;,
\label{Tmod}
\ee
where $\phi_{10}$ is the ten-dimensional dilaton, $g_{i\bar i }$ is 
the metric component along the $T^2_i$ torus and $\theta_i$ is
a RR axion. The real part of these moduli is therefore represented 
by a NSNS vertex operator, whereas the imaginary part is described by 
a RR vertex, involving spin-fields and particularly unpleasant to deal 
with. Notice for the moment that these vertex operators can provide 
at most one spacetime fermionic zero-mode. Since physical gluons and 
gravitons bring each two fermionic zero-modes, correlations with an 
odd number of moduli vanish, as expected from the fact these should 
come in pairs reconstructing composite connections. Moreover, in the 
limit of interest, the correlation functions under analysis factorize 
into an internal correlation among moduli fields and a spacetime correlation 
among gluons and gravitons.

We now propose an approach to the derivation of the effective vertex
for the composite connection, which is not exhaustive but will allow
us to emphasize a few important points. Focus for simplicity on a 
single internal torus only, for which the composite curvature 
(\ref{FKex}) becomes (no sum over the indices) 
$G_{i,\mu\nu} = 2i \, K_{i\bar i}\, 
\partial_{[\mu}t^i \partial_{\nu]} \bar t^i$, with
$K_{i\bar i}=(t^i+\bar t^i)^{-2}$.
On general grounds, one expects the moduli to pair and reconstruct only 
composite curvatures of this form. At leading order in the momenta,
the structure of the internal correlation between two moduli must 
therefore be as follows:
\bea
&& \langle V_{t^i}(p_1) V_{\bar t^i}(p_2) \rangle 
= \alpha_i \, K_{i \bar i} \, p_{1\mu} t^i \, p_{2\nu} \bar t^i \, 
\psi_0^\mu \, \psi_0^\nu \label{4pt1} \;, \\
&& \langle V_{t^i} V_{t^i}\rangle 
= \langle V_{\bar t^i} V_{\bar t^i}\rangle=0 \label{4pt2} \;,
\eea
where $\alpha_i$ are some coefficients and $V_{t^i}$ and
$V_{\bar t^i}$ are the vertex operators for the scalars $t^i$ and
$\bar t^i$. As already mentioned, correlations 
such as (\ref{4pt1}) are potentially difficult to compute in 
orientifold models, because the moduli vertices have a simple NSNS 
real part, but a complicated RR imaginary part. 
More precisely, the sigma-model curvature can be rewritten as 
$G_{i\mu\nu}= i\,K_{i\bar i}\, \partial_{[\mu} (t^i - \bar t^i) 
\partial_{\nu]} (t^i + \bar t^i)$, and one has in principle to use one 
RR vertex $V_{t^i}-V_{\bar t^i}$ and one NSNS vertex 
$V_{t^i}+V_{\bar t^i}$. 
One could then proceed by contracting the NSNS and RR vertex, take
the $\alpha^\prime \rightarrow 0$ limit and try to figure out which is the
effective vertex that, inserted in the correlation function, gives
the same result. This procedure is however complicated, so we prefer
to use a trick that will allow us to deduce the effective vertex
in a quicker (although not rigorous) way.

The point is that correlations involving only pairs of $V_{t^i}+V_{\bar t^i}$
vertices are formally proportional to the corresponding correlations
involving pairs of $V_{t^i}+V_{\bar t^i}$ and $V_{t^i}-V_{\bar t^i}$
vertices. Indeed, using (\ref{4pt1}) and (\ref{4pt2}), one gets:
\be
\langle (V_{t^i} \pm V_{\bar t^i})(p_1) 
(V_{t^i}+ V_{\bar t^i})(p_2) \rangle \nn \\ 
= \alpha_i \, K_{i \bar i}\,\Big(p_{1\mu} t^i \, p_{2\nu} \bar t^i 
\pm (1 \leftrightarrow 2) \Big) \psi_0^\mu \psi_0^\nu \;. \label{4pt3}
\ee
Due to the symmetrization in $1 \leftrightarrow 2$, one 
gets a vanishing result for two NSNS vertices (upper sign), but a non 
vanishing one for one RR and one NSNS vertices (lower sign). 
Nevertheless, both of them encode the same non-vanishing coupling 
$\alpha_i$, and by careful inspection it is possible to extract the latter 
also from the vanishing correlation involving only NSNS vertices, after 
having recognized the zero corresponding to the unavoidable symmetrization. 
A convenient way to properly remove the zero is to flip the crucial sign 
by hand in the final result, reconstructing the sigma-model curvature.
A similar analysis goes through for correlations involving more than two 
moduli. Indeed, as will now become clear, the moduli vertices do indeed 
always contract in pairs associated to composite curvatures, and all of 
them can be represented by the NSNS real part, keeping track of the zeroes 
arising by symmetrization. 

We are now in position to attempt a derivation of the effective vertex 
operator for the composite connections (\ref{FKexc}), by considering a 
correlation involving an even number of moduli real parts and using 
the trick discussed above. The corresponding NSNS vertex operator 
can be easily deduced from (\ref{Tmod}), and is given by
\be
V_{t^i}+V_{\bar t^i}= (t^i + \bar t^i) \int\! d^2z \, 
(\partial X^i+ip\cdot\psi\psi^i)\,
(\bar{\partial}\bar{X}^i+ip\cdot\tilde\psi\bar{\tilde\psi^i}) 
\, e^{ip\cdot X} + \;\; {\rm c.c.} \;,
\label{ReT}
\ee
where ${\rm c.c.}$ stands for complex conjugate\footnote{Notice that 
the vertex (\ref{ReT}) is actually the right one for $g_{i\bar i}$, 
that differs from $t^i+\bar t^i$ for a factor $g_S=e^{-\phi_{10}}$. 
This difference, possibly important for a careful understanding and 
comparison of string and field theory results (see e.g. footnote 12),
is however irrelevant for most of the considerations that will follow. 
Correspondingly, we effectively identify $g_{i\bar i}$ with $t^i+\bar t^i$.}.
This vertex can be further simplified case by case thanks to the limit 
$\alpha^\prime \rightarrow 0$, and to the presence of fermionic 
zero-modes. But contrarily to the simpler case of gluons and 
gravitons, it might happen that pieces of the vertex which are
apparently subleading for small momenta,
give nevertheless a leading contribution when contracted.
We proceed separately for the $A$, $M$, $K$ and the 
$T$ surfaces.

\vskip 9pt
\noindent
{\bf $A$, $M$ and $K$ surfaces}
\vskip 3pt
\noindent
In this case, one can start with the following effective vertex:
\be
V_{\bar t^i}+V_{t^i} = ip\cdot\psi_0 \, (t^i + \bar t^i) \int\! d^2z \, 
\Big[\psi^i \bar \partial \bar X^i 
+ \tilde \psi^i \partial \bar X^i
+ \bar \psi^i \bar \partial X^i 
+ \bar {\tilde \psi^i} \partial X^i + ...\Big] \;,
\label{vtteff}
\ee
where the dots represent possibly important fermionic terms, that are 
difficult to fix unambiguously in the present approach. 
By exponentiating two of these vertices with momentum $p^i_{1,2}$, and
performing a shift on the internal fermions, one gets an effective 
interaction for the internal bosons. Rescaling then
$(X^i,\bar X^i)\rightarrow g_{i\bar i}^{-1/2}(X^i,\bar X^i)$ so that
the bosonic kinetic terms are normalized, one finds
\be
S_{int} = K_{i\bar i} \Big(p_{1\mu} t^i \, p_{2\nu} \bar t^i 
+ (1 \leftrightarrow 2) \Big) \psi_0^\mu \psi_0^\nu \int\! d^2z \, 
\Big[\bar X^i (\partial + \bar \partial) X^i + ... \Big] \;.
\label{Sint1}
\ee
As expected, the factor in front of the effective vertex (\ref{Sint1})
has precisely the same form as (\ref{4pt3}) with the $+$ sign,
and this interaction term vanishes due to the 
$1 \leftrightarrow 2$ symmetrization. According to the previous discussion, 
by flipping the sign of the second term in the brackets, one generates 
a non-vanishing interaction which 
can be interpreted as an effective vertex operator for the composite 
connection. Notice that one would expect such an effective 
vertex to be world-sheet supersymmetric, whereas the expression obtained 
above is not. We conclude from this that the expression (\ref{Sint1}) is  
incomplete, and that additional purely fermionic terms must indeed be present 
in (\ref{vtteff}) and (\ref{Sint1}). 
By requiring a world-sheet supersymmetric vertex, it is
then easy to deduce the right form for these fermionic terms, and one 
finds finally 
\be
V^{eff.}_G =  \frac 12\, G_{i,\mu\nu} \psi_0^\mu \psi_0^\nu \, 
\int\! d^2z \, \Big[\bar X^i (\partial + \bar \partial) X^i 
+ (\bar \psi - \bar{\tilde \psi})^i (\psi - \tilde \psi)^i \Big] \;.
\label{VKeff}
\ee

\vskip 9pt
\noindent
{\bf $T$ surface}
\vskip 3pt
\noindent
In this case, on can effectively take:
\be
V_{\bar t^i}+V_{t^i} = ip\cdot\psi_0 \,(t^i + \bar t^i) 
\int\! d^2z \, \Big[\psi^i \partial \bar X^i 
+ \bar \psi^i \partial X^i + ... \Big] \;.
\ee
The dots represent again possible fermionic terms. 
By exponentiating and performing a shift on the left-moving internal 
fermions, one gets an effective interaction for the bosons given by
\be
S_{int} = K_{i\bar i}\Big(p_{1\mu} t^i \, p_{2\nu} 
\bar t^i + (1 \leftrightarrow 2) \Big) 
\psi_0^\mu \psi_0^\nu \int\! d^2z \, 
\Big[\bar X^i \bar \partial X^i + ... \Big] \;.
\label{Sint2}
\ee
As before, this interaction term vanishes and one has to perform the 
discussed sign flip to obtain a non-vanishing interaction to be
interpreted as an effective vertex operator for the composite connection.
Again, since such effective vertex should be world-sheet supersymmetric,
we conclude that (\ref{Sint2}) is indeed incomplete, and
fix again the missing fermionic terms thanks to world-sheet
supersymmetry. Finally, one gets
\be
V^{eff.}_G =  \frac 12\, G_{i,\mu\nu} \psi_0^\mu \psi_0^\nu \, 
\int\! d^2z \, \Big[\bar X^i \bar \partial X^i
+ \bar{\tilde \psi^i} \tilde \psi^i \Big]\;.
\label{VKeff2}
\ee

\vskip 10pt

There is an alternative way to deduce the form of the effective 
vertices above. Since the NSNS ${\rm Re}\,t^i$ scalar is related to
the metric of the corresponding internal two-torus, the exponentiation
of its vertex induces a geometric deformation of the orbifold
along the $i$-th internal torus. This can be analyzed directly from a 
$\sigma$-model point of view. By doing that, with standard techniques,
it is easy to see that the metric deformation associated to the internal
$T^2_i$ torus is represented by (\ref{VKeff}) and (\ref{VKeff2}) on the
corresponding surfaces, where $G_i$ is now replaced by the geometric
curvature of $T^2_i$. By exploiting the tensorial structure of this 
curvature, one easily realizes that the components whose derivatives
are all along the spacetime directions, like in (\ref{Sint1}), vanish
due to a symmetrization, exactly like before. As expected, one is 
therefore led to use the same trick as above to get a non-vanishing
composite field-strength. However, in this way one gets automatically 
the fermionic terms in (\ref{VKeff}) and (\ref{VKeff2}) and also a 
first clue of the close relation between the field-strength $G$ and 
the curvature of the internal space. We postpone to Section 5 a more
precise analysis of this relationship.

Notice also that in heterotic models, where the untwisted moduli
consist of NSNS fields only, the correspondence between K\"ahler
deformations of the orbifold and sigma-model symmetry can be unambiguously
established. The net result is again that the effective vertex for the 
composite connection has the same form as that of an internal graviton,
like in (\ref{VKeff2}).
We think that this gives some extra evidence for the relation between 
sigma-model symmetry and orbifold K\"ahler deformations also in Type IIB 
orientifolds. Indeed, although in the latter case the pseudo-scalars
${\rm Im}\,t^i$ are RR fields, from a purely geometrical point of
view there is no difference with respect to heterotic models,
since in both theories ${\rm Im}\,t^i$ simply complexifies the 
geometric K\"ahler structure of the orbifold/orientifold.


\section{String computation} 

The computation of the partition functions entering the anomaly polynomial
closely follow \cite{ss2,ss3}. We proceed separately for the various
surfaces. The $A$, $M$ and $K$ amplitudes are generalizations of the 
results of \cite{ss2,ss3} to a non-trivial ``composite'' background. 
The $T$ amplitude was instead irrelevant in \cite{ss2,ss3}, as shown 
in Appendix B for the six-dimensional case, and has therefore 
to be computed in detail.

As already said, we restrict to the simplest $\Z_3$ and $\Z_7$ models, 
which do not contain D5-branes neither $N=2$ sub-sectors. In these models, 
the $k$-th element of $\Z_N$ is $g^k = (\theta^k,\gamma_k)$, where $\theta^k$ 
is a rotation of angles $2 \pi k v_i$ in the internal two-tori $i=1,2,3$, 
and $\gamma_k$ is a non-trivial twist matrix, acting on the Chan-Paton bundle. 
The Chan-Paton representation of the twist is fixed by the
tadpole cancellation condition. For future convenience, and in order to 
get contact with the notation used in the literature, we define
\be
C_k = \prod_{i=1}^3 (2\sin \pi kv_i) \;.
\label{Ck}
\ee
and its sign $\epsilon_k = {\rm sign}\, C_k$. For $N$ odd, the tadpole 
cancellation condition can then be written as 
\be
\frac 14 {\rm tr}(\gamma_{2k}) = \frac {C_{2k}}{C_k}
= \frac {C_{k}}{C_{2k}}
\label{tadpole}
\ee
and holds because actually all the quantities in the equality 
are equal to a sign, namely $\epsilon_{2k}/\epsilon_k$ which is 
equal to $-1$ for $\Z_3$ and $+1$ for $\Z_7$.

Let us define the characteristic classes which will 
appear in the polynomial associated to generic 
sigma-gauge-gravitational anomalies. 
For the gauge bundle, one has the natural $\Z_N$ Chern character,
function of the gauge curvature $F$, defined as a trace over the 
Chan-Paton representation:
\be 
{\rm ch}_{k} (F) = {\rm tr} \, [\gamma_k \,e^{i F/2\pi}] \;.
\label{chern}
\ee 
This factor appears in the anomaly from charged chiral spinors.
For the tangent bundle, the relevant characteristic classes
are the Roof-genus, G-polynomial and Hirzebruch polynomial,
functions of the gravitational curvature $R$ and defined in terms 
of the skew eigenvalues $\lambda_a$ of $R$ as: 
\bea
&& \widehat{A}(R) = \prod_{a=1}^{D/2} \frac{\lambda_a/4\pi}
{\sinh \lambda_a/4\pi} \;, \label{Ahat} \\ 
&& \widehat{G}(R) = \prod_{a=1}^{D/2} \frac{\lambda_a/4\pi}
{\sinh \lambda_a/4\pi}\Big(2 \sum_{b=1}^{D/2} 
\cosh \lambda_b/2\pi \,-\, 1\Big) \;, \label{Ghat} \\
&& \widehat{L}(R) = \prod_{a=1}^{D/2} \frac{\lambda_a/2\pi}
{\tanh \lambda_a/2\pi} 
\label{Lhat} \;.
\eea
These factors appear respectively in the anomaly from chiral spinors,
chiral Rarita-Schwinger fields, and self-dual tensor fields.
We also introduce three new characteristic classes depending on 
the composite curvature $G=dZ$, defined in terms
of the curvatures $G_i$ in the three internal tori as
\bea
\widehat A_k(G) &=& \prod_{i=1}^3
\frac {\sin (\pi k v_i)}{\sin (\pi k v_i + G_i/2\pi)} \;, 
\label{Ak} \\
\widehat G_k(G) &=& \prod_{i=1}^3
\frac {\sin (\pi k v_i)}{\sin (\pi k v_i + G_i/2\pi)} 
\Big(2 \sum_{j=1}^{3} \cos (2 \pi k v_j + G_j/\pi) \,-\, 1\Big) \;, 
\label{Gk} \\
\widehat L_k(G) &=& \prod_{i=1}^3
\frac {\tan (\pi k v_i)}{\tan (\pi k v_i + G_i/\pi)} \;.
\label{Lk}
\eea
These characteristic classes will appear in the anomaly from states 
transforming as chiral spinors, Rarita-Schwinger fields and self-dual 
tensors with respect to sigma-model transformations.

A last preliminary comment relevant to all the surfaces is the following. 
Due to the universal six bosonic zero-modes in the four non-compact spacetime
directions and the two extra auxiliary dimensions introduced to deal with the 
WZ descent, the partition functions will always contain a free-particle 
contribution proportional to $\rho^{-3}$. Moreover, the curvatures will
always appear multiplied by $\rho$ as twists in the partition function. 
An important simplification occurs using the fact that only the 6-form
component of the partition function is relevant for our purposes: one 
can scale out the above explicit dependences on the modulus $\rho$.
This will be important also for the modular invariance of the string 
amplitudes yielding the anomaly in the torus surface, as we shall see.

\subsection{\bf $A$, $M$ and $K$ surfaces}

On the $A$, $M$ and $K$ surfaces, the boundary of moduli space
is given by the component $t \rightarrow \infty$ encoding the quantum
anomaly, minus the component $t \rightarrow 0$ encoding the classical 
GS inflow. The contribution of each surface to the total anomaly
polynomial is given by
\be
I_{\Sigma} = \Big(\lim_{t \rightarrow \infty} - \lim_{t \rightarrow 0} \Big) 
Z_\Sigma (t) \;.
\label{IAMK}
\ee
The partition functions $Z_\Sigma (t)$ are in the RR odd spin-structure,
and their operatorial representation is
\bea
&& Z_{A}(t) = \frac{1}{4N}\sum_{k=0}^{N-1}\,
{\rm Tr}_{R}\,[g^k\,(-1)^F\,e^{-tH}] \;,\nn \\ 
&& Z_{M}(t) =\frac{1}{4N}\sum_{k=0}^{N-1}\, 
{\rm Tr}_R\,[\Omega\,g^k\,(-1)^F\,e^{-tH}] \;,\nn \\
&& Z_{K}(t) =\frac{1}{8N}\sum_{k=0}^{N-1}\, 
{\rm Tr}_{RR}\,[\Omega\,g^k\,(-1)^{F+\tilde F}\,e^{-tH}] \;.
\label{AMK} 
\eea
Here $H=H(R,F,G)$ is the Hamiltonian associated to the two-dimensional 
supersymmetric non-linear $\sigma$-model in a gauge, gravitational and 
composite background defined by the effective vertex operators below, with Neumann 
boundary conditions. Due to supersymmetry, (\ref{AMK}) are generalized 
Witten indices in which only massless modes can contribute \cite{wit}. 
Indeed, it can be verified explicitly that massive world-sheet fermionic 
and bosonic modes exactly cancel. As a consequence, the partition functions 
(\ref{AMK}) are independent of $t$, and (\ref{IAMK}) vanishes, reflecting 
anomaly cancellation through the GS mechanism.

The background dependence of the action is encoded in the effective vertices
for external particles. In the odd spin-structure on the $A$, $M$ and $K$
surfaces, the sum $Q + \tilde Q$ of the left and right world-sheet 
supersymmetries is preserved, and there are space-time fermionic zero-modes 
$\psi_0^\mu = \tilde \psi_0^\mu$. In the limit $\alpha^\prime \rightarrow 0$, 
we use the following effective vertex operators for gluons, gravitons
and composite sigma-model connections:
\bea
&& V^{eff.}_F = F^a\,\oint\!d\tau \lambda^a\;,\\
&& V^{eff.}_R = R_{\mu\nu}\, \int\!d^2z\, 
\left[X^\mu (\partial + \bar \partial) X^\nu + 
(\psi - \tilde \psi)^\mu (\psi - \tilde \psi)^\nu \right] \;,\\
&& V^{eff.}_G =  G_i \, \int\! d^2z \, 
\Big[\bar X^i (\partial + \bar \partial) X^i 
+ (\bar \psi - \bar{\tilde \psi})^i (\psi - \tilde \psi)^i \Big] \;,
\eea
in terms of the curvature two-forms
\be
F^a = \frac 12\, F_{\mu\nu}^a \, \psi_0^\mu \psi_0^\nu \;,\;\;
R_{\mu\nu} = \frac 12\, R_{\mu\nu\rho\sigma}\, \psi_0^\rho \psi_0^\sigma 
\;,\;\; G_i = \frac 12\, G_{i,\mu\nu} \psi_0^\mu \psi_0^\nu \;.
\label{curvggk}
\ee
It is now straightforward to compute the partition functions (\ref{AMK}) 
on the $A$, $M$ and $K$ surfaces. The composite background modifies only 
the internal partition functions, whereas the spacetime contribution 
has only the standard dependence on the gauge and gravitational backgrounds. 
The spacetime part can be computed exactly as in \cite{mss,ss1}, and one 
finds the same results as in \cite{ss2,ss3}. The computation of the internal 
part is also similar to that in \cite{ss2,ss3}, the curvature $G$ 
entering as a twist. Using $\zeta$-function regularization, one finds
\bea
&& Z_{A} = \frac{i}{4N}\,\sum_{k=1}^{N-1}
C_k\, \widehat A_k(G)\,{\rm ch}_k^2 (F)\,
\widehat A(R) \;, \nn \\
&& Z_{M} = - \frac{i}{4N}\,\sum_{k=1}^{N-1}
C_k\, \widehat A_k(G)\, {\rm ch}_{2k}(2F)\, 
\widehat A(R) \;, \nn \\
&& Z_{K} = \frac{i}{16N}\,\sum_{k=1}^{N-1}
C_{2k}\, \widehat L_k(G)\, 
\widehat L(R) \;, \label{amkkahler}
\eea
in terms of the characteristic classes defined before.
As anticipated, the partition functions (\ref{amkkahler}) are independent 
of the modulus $t$. Consequently, the quantum anomaly encoded in the 
$t \rightarrow \infty$ boundary, and the classical inflow associated to 
$t \rightarrow 0$ boundary, are precisely opposite to each other and cancel 
on each of the $A$, $M$ and $T$ surfaces.

\subsection{\bf $T$ surface}

On the $T$ surface, the boundary ${\partial \cal F}$ of 
moduli space splits into the component at infinity, 
${\partial \cal F}_{\infty} = [-1/2 + i\,\infty, 1/2 + i\,\infty]$,
minus the remaining component, ${\partial \cal F}_0$, and the 
contribution to the total anomaly polynomial is given by 
\be
I_{T}= \frac 12 \left[\left( \oint_{{\partial \cal F}_\infty} 
\!\!\!\! - \oint_{{\partial \cal F}_0}\! \right) d\tau \,Z_T (\tau) 
+ \left( \oint_{{\partial \cal F}_\infty} 
\!\!\!\! - \oint_{{\partial \cal F}_0}\!\right) d\bar \tau\, Z_T (\bar \tau) 
\right] \;.
\label{anotoro}
\ee
The quantities
\be
Z_T (\tau) = \sum_{\alpha} (-1)^{\alpha} Z_T^{S_\alpha}(\tau) \;,\;\;
Z_T (\bar \tau) = \sum_{\tilde\alpha} 
(-1)^{\tilde \alpha} Z_T^{S_{\tilde\alpha}}(\bar \tau) \;.
\label{part}
\ee
are the total partition functions in the odd-even and even-odd sector
respectively. More precisely, $\alpha = 2,3,4$ represent the RR, 
${\rm RNS}_+$ and ${\rm RNS}_-$ odd-even spin-structures, and similarly 
$\tilde \alpha = 2,3,4$ represent the RR, ${\rm RNS}_+$ and 
${\rm RNS}_-$ even-odd spin-structures. 
Their operatorial representation is
\bea
&& Z_{T}^{RR}(\tau) = \frac{1}{8N}\sum_{k,l=0}^{N-1}\,
{\rm Tr}^{(l)}_{RR}\,[g^k\,(-1)^F\,\tilde g^k\,
e^{-\tau H}\,e^{- \bar \tau \tilde H}] \;,\nn \\ 
&& Z_{T}^{RNS_+}(\tau) = \frac{1}{8N}\sum_{k,l=0}^{N-1}\,
{\rm Tr}^{(l)}_{RNS}\,[g^k\,(-1)^F\,\tilde g^k\,
e^{-\tau H}\,e^{- \bar \tau \tilde H}] \;,\nn \\ 
&& Z_{T}^{RNS_-}(\tau) = \frac{1}{8N}\sum_{k,l=0}^{N-1}\,
{\rm Tr}^{(l)}_{RNS}\,[g^k\,(-1)^F\,\tilde g^k\,(-1)^{\tilde F}\,
e^{-\tau H}\,e^{- \bar \tau \tilde H}] \:.
\label{T234}
\eea
The expression for the even-odd spin-structures is perfectly similar,
with left and right movers exchanged.
In this case, $H$ and $\tilde H = \tilde H(R,F,G)$ are the 
left and right-moving Hamiltonians associated to the two-dimensional 
supersymmetric non-linear $\sigma$-model in a gauge, gravitational 
and composite background defined by the effective vertex operators below.
Notice that whereas the even part of the partition functions is influenced
by the backgrounds, the odd part remains trivial. This will lead to
holomorphic and anti-holomorphic results in the odd-even and even-odd
spin structures. Furthermore, only the odd parts of (\ref{T234}) are 
supersymmetric indices, whereas the even parts receive contributions 
from all the tower of string states and will therefore depend on $\tau$.

Again, the background dependence of the action is encoded in the effective 
vertices for external particles. In the odd-even spin-structure on the $T$
surface, the left-moving world-sheet supersymmetry $Q$ is preserved, and 
there are space-time fermionic zero-modes $\psi_0^\mu$. In the limit 
$\alpha^\prime \rightarrow 0$, we use the following effective vertex 
operators for gravitons and composite connections:
\bea
&& V^{eff.}_R = R_{\mu\nu}\, \int\!d^2z 
\Big[X^\mu \bar \partial X^\nu + \tilde \psi^\mu \tilde \psi^\nu \Big] \;,
\label{gravtor} \\
&& V^{eff.}_G =  G_i \, \int\! d^2z \, 
\Big[\bar X^i \bar \partial X^i + \bar{\tilde \psi^i} \tilde \psi^i \Big]\;,
\label{kator}
\eea
in terms of the curvature two-forms defined in (\ref{curvggk}).
It is then easy to evaluate the partition function on the $T$ surface. 
The gravitational background influences bosons and left-moving 
fermions, in a similar way to the cases discussed in Appendix B.
The composite background influences instead only the internal bosons
and left-moving fermions. 
The evaluation of the internal partition functions is very similar
to that reported in Appendix B for the six-dimensional case of 
Type IIB on $T^4/\Z_N$, the curvature $G$ being responsible
for a twist. In total, one gets:
\bea
Z_T (R,G,\tau) &=& \frac {i}{8N} \sum_{\alpha=2}^4 (-1)^\alpha
\sum_{k,l=0}^{N-1} N_{k,l} \prod_{i=1}^{3} 
\frac{\theta_\alpha{lv_i \brack kv_i}(-G_i/\pi^2|\tau)}
{\theta_1{lv_i \brack kv_i}(-G_i/\pi^2|\tau)} \nn \\ 
&\;& \hspace{95pt} \times \prod_{a=1}^2 
\left[\frac{(i x_a )}{\theta_1(ix_a/\pi|\tau)} 
\,\theta_\alpha(i x_a/\pi|\tau) \right] 
\frac{\eta^3(\tau)}{\theta_\alpha(0|\tau)} \;, \hspace{25pt}
\label{T}
\eea
where $x_a=\lambda_a/2\pi$ and $N_{k,l}$ is the number of fixed-points 
that are at the same time $k$ and $l$-fixed ($N_{0,0} = 0$).
The result for the odd-even spin-structures is the complex conjugate of
(\ref{T}). 

It is a lengthy but straightforward exercise to show that the partition 
function (\ref{T}) is modular invariant. Indeed, one gets
\bea
Z_T(R,G,\tau +1) &=& Z_T (R,G,\tau) \nn \\
Z_T (R,G,-1/\tau) &=& \frac 1\tau Z_T (R\tau,G\tau,\tau)
= \tau^2 Z_T (R,G,\tau) \;,
\eea
where the last step in the second equation is valid for the relevant 6-form 
component of $Z_T$. Thanks to the modular invariance of 
$Z_T (\tau)$ and $Z_T (\bar \tau)$, their integral on various components
of ${\partial \cal F}$ are related to each other. In fact, only
the component $\partial {\cal F}_\infty$ at infinity can give a 
non-vanishing contributions, the remaining four pieces of the remaining 
component $\partial {\cal F}_0$ canceling pairwise, as in \cite{susu}.
The potential contribution from $\partial {\cal F}_\infty$ is interpreted 
as a quantum sigma-gravitational anomaly.

In order to evaluate the contribution from $\partial {\cal F}_\infty$,
one has to take the limit $\tau_2\rightarrow\infty$ of the partition 
function. This is easy to take in untwisted sectors, but in twisted 
sectors one has to pay attention to the range of the twists. 
For $l\neq 0$, one gets for instance:
\bea
&& \prod_{i=1}^{3} \frac{\theta_2{lv_i \brack kv_i}(-G_i/\pi^2|\tau)}
{\theta_1{lv_i \brack kv_i}(-G_i/\pi^2|\tau)} \rightarrow 
-\,i\,\epsilon_l \;, \nn \\ && \prod_{i=1}^{3} 
\frac{\theta_{3,4}{lv_i \brack kv_i}(-G_i/\pi^2|\tau)}
{\theta_1{lv_i \brack kv_i}(-G_i/\pi^2|\tau)} \rightarrow 
\mp \, i \, \epsilon_l \, q^{1/8} \prod_{i=1}^{3} 
\exp(-\,i\,\epsilon_l G_i/\pi) \;. \nn
\eea
where the quantity $\epsilon_k$ was defined as the sign of 
$C_k$ in (\ref{Ck})\footnote{It arises here as 
$\epsilon_l = (-1)^{\sum_i \theta_i (l v_i)}$ and  
$\epsilon_l = 2 \sum_i \theta_i (l v_i) - 3$ in terms of the 
representative $\theta_i (l v_i) = l v_i - {\rm int}(l v_i)$
of the twist $l v_i$ in the interval $[0,1]$.}.
The above expressions already show that the anomaly from RR twisted 
states does not depend on the composite curvature, and therefore trivially 
vanishes in $D=4$. On the contrary, the anomaly from RNS twisted states
does depend on the curvature $G$, and is non-vanishing. 
The corresponding apparently complex internal contribution to the 
partition function (\ref{T}) turns out to be actually real, and
using the fact that only odd powers of $G_i$ 
are relevant in $D=4$, one gets:
$$
\mp \, i \, \epsilon_l \, q^{1/8} \prod_{i=1}^{3} 
\exp(-\,i\,\epsilon_l G_i/\pi) 
= \mp \, q^{1/8} \, (-i)\, {\rm ch}\,(2 G)\;.
$$

Finally, the total odd-even and even-odd spin structure partition 
functions (\ref{part}) are found to behave both in the same way
in the limit $\tau_2 \rightarrow \infty$, giving:
\bea
Z_T &\rightarrow& -\frac i{16N} \sum_{k=1}^{N-1}
C_{2k} \, \widehat L_k(G) \, \widehat L(R) \nn \\ 
&\;& + \frac {i}{2N} \sum_{k=1}^{N-1}
C_k \left[\widehat A_k (G) \, \widehat G(R) 
+ \widehat G_k (G)\, \widehat A(R)\right] \nn \\ &\;& 
+ \frac {i}{2N} \sum_{k=0}^{N-1} \sum_{l=1}^{N-1} N_{k,l} \, 
(-i)\, {\rm ch}\,(G) \,\widehat A(R) \;. \label{Ttot}
\eea
Since this expression is independent of $\tau$, the remaining 
integral over $\tau_1$ in $\partial {\cal F}_\infty$ is trivial,
and according to (\ref{anotoro}), this is also the final result for 
the $T$ contribution to the anomaly.
The first line of (\ref{Ttot}) corresponds to the RR untwisted sector,
whose contribution precisely cancels that of the Klein bottle in 
(\ref{amkkahler}). The second line corresponds to the RNS/NSR 
untwisted sectors and encodes the contributions of the gravitino, 
dilatino and untwisted modulini. Finally, the last line encodes those 
of twisted RNS/NSR moduli; notice that all the twisted sectors 
$l=1,...,N-1$ give the same contribution, since $N_{k,l}$ takes the 
same value for all $\{k,l\} \neq \{0,0\}$ for $N$ odd.
Actually, one can check that the relevant 6-form component of the 
result (\ref{Ttot}) vanishes identically. Some useful details 
in this respect are reported in Appendix C. In conclusion, the total
anomaly from the $T$ surface exactly cancels:
\be
Z_T \rightarrow  0 \;.
\ee

Note that whereas the vanishing of the $T$ amplitude is expected from
modular invariance in a full string context, it has to be explicitly 
checked in the particular $\alpha^\prime \rightarrow 0$ limit we consider.
Because of the importance of this result and since we are not aware of 
any similar computation in the literature, we report in Appendix B a 
similar computation of gravitational anomalies on the $T$  surface
for Type IIB string theory in $D=10$ and $D=6$ on an orbifold.


\section{Topological interpretation}

Probably it is interesting to point out that all the anomalies considered
so far, eqs.(\ref{amkkahler}), have a nice topological interpretation in terms
of the $G$-index of the Dirac operator ($A$ and $M$)
and the $G$-index of the signature complex \cite{ASg1,ASg2} ($K$), 
with $G=\Z_N$ for a $\Z_N$ orbifold (see \cite{sha} for a nice 
introduction and more details on the $G$-index)\footnote{A relation 
between anomalous couplings and the $\Z_2$ signature complex was 
already exploited in \cite{ss1} in the case of smooth manifolds.}.

The $\Z_N$ group can be thought to act on the whole ten dimensional
spacetime $X=R^{1,3}\times T^6$, as well as on the gauge bundle. 
As before, we denote by $g_k=(\theta^k,\gamma_k)$ the $k$-th element 
of the complete $\Z_N$ group. Among other things, this will twist
the Chern classes appearing in index theorems. The subspace $X_k$ 
left invariant by the geometric $\theta^k$ is 
$X_k= \oplus_{i=1}^{N_k} R^{1,3}$, that is $N_k$ copies of spacetime. 
When restricted to $X_k$, the tangent bundle of $X$ decomposes into the 
tangent and normal bundles ${\cal T}_k$ and ${\cal N}_k$ of $X_k$ in $X$. 
Moreover, the normal bundle ${\cal N}_k$ further decomposes naturally 
into three components ${\cal N}^i_k$, in which $\theta$ acts as $2\pi v_i$ 
rotations. The cotangent and spin bundles, which will be relevant for 
spinor and self-dual tensor fields, have a similar decomposition. 

The Dirac-$G$ index theorem is then given by (see e.g. \cite{sha})
\be
\mbox{index}(D_{g^k})=\int_{X^G}\, \frac{
{\rm ch} (S^+_{{\cal T}_k} - S^-_{{\cal T}_k})\,
{\rm ch}_k (S^+_{{\cal N}_k} - S^-_{{\cal N}_k})\,{\rm ch}_k (F)}
{{\rm ch}_k(\tilde{\cal N}_k)\,e({\cal T}_k)} \,{\rm Td}({\cal T}_k^{C})
\label{Dg}
\ee
where $S^\pm_{{\cal T}_k}$ and $S^\pm_{{\cal N}_k}$ are the positive 
and negative chirality spin bundles lifted from the tangent and normal 
bundles, and $\tilde{\cal N}_k=\oplus_i (-)^i \wedge^i {\cal N}^*_k$ in 
terms of the conormal bundle ${\cal N}^*_k$. $e({\cal T}_k)$ and 
${\rm Td}({\cal T}_k^{C})$ are the usual Euler and (complexified)
Todd classes:
$$
{\rm Td}({\cal T}_k^{C}) = \prod_{a=1}^{2} \frac{x_a}{1-e^{-x_a}}
\frac{(-x_a)}{1-e^{x_a}} \;,\;\; e({\cal T}_k)=\prod_{a=1}^2 x_a \;. 
$$
By expliciting the other terms appearing in (\ref{Dg}), one gets
\bea
&& {\rm ch} (S^+_{{\cal T}_k} - S^-_{{\cal T}_k}) 
= \prod_{a=1}^{2} \, (e^{x_a/2}-e^{-x_a/2}) \;, \nn \\
&& {\rm ch}_k (S^+_{{\cal N}_k} - S^-_{{\cal N}_k}) 
= \prod_{i=1}^3 \,(e^{x_i/2}e^{i\pi kv_i}-e^{-x_i/2}e^{-i\pi kv_i}) \;, \nn \\ 
&& {\rm ch}_k(\tilde{\cal N}_k) = \prod_{i=1}^{3} \, 
(1-e^{x_i}e^{2i\pi kv_i})\,(1-e^{-x_i}e^{-2i\pi kv_i}) \;,
\label{esp}
\eea
where $x_a$ and $x_i$ are the eigenvalues of the curvature two-form 
on ${\cal T}_k$ and ${\cal N}_k$. ${\rm ch}_k(F)$ is precisely the
twisted Chern character defined in (\ref{chern}), in terms of the twist 
matrix $\gamma_k$. The trace is in the bifundamental or fundamental 
representation of the gauge group, for the $A$ and $M$ surfaces respectively.
As previously discussed, the composite field-strength $G$ is closely
associated to the curvature two-form of the normal bundle ${\cal N}_k$.
More precisely, $x_i=i\,G_i/\pi$, and by plugging in the relations 
(\ref{esp}) above, one gets after some simple algebra
\be
\mbox{index}(D_{g^k})=-i \int_{R^{3,1}} C_k \, \, \widehat A_k(G) 
\, {\rm ch}_k(F) \, \widehat A(R) \;,
\ee
which corresponds to the k-th term in the partition functions 
(\ref{amkkahler}) on $A$ and $M$.

The case of the $G$-index of the signature complex can be treated
similarly. The $G$-signature index theorem is 
\be
\mbox{index}({\cal D}^+_{g^k})=\int_{X^G}\, \frac{{\rm ch} 
({\cal T}_k^+-{\cal T}_k^-)\,{\rm ch}_k({\cal N}_k^+-{\cal N}_k^-)}
{{\rm ch}_k(\tilde{\cal N}_k)\,e({\cal T}_k)} {\rm Td}({\cal T}_k^{C})
\ee
where ${\cal T}_k^{\pm}={}^{\pm}\!\!\wedge {\cal T}^*_k$, 
${\cal N}_k^{\pm}={}^{\pm}\!\!\wedge {\cal N}^*_k$, in terms of the 
cotangent and conormal bundles ${\cal T}^*_k$ and ${\cal N}^*_k$.
More explicitly, we have
\bea
&& {\rm ch} ({\cal T}_k^+-{\cal T}_k^-) = \prod_{a=1}^{2} \, 
(e^{x_a}-e^{-x_a}) \;, \nn \\
&& {\rm ch}_{k}({\cal N}_k^+-{\cal N}_k^-) = \prod_{i=1}^{3} \, 
(e^{-x_i}e^{-2i\pi k v_i} - e^{x_i}e^{2i\pi kv_i}) \;.
\eea
Similarly to the previous case, the index can then be written as 
\be
\mbox{index}({\cal D}_{g^k}^+)=-i\int_{R^{3,1}} C_{2k} \, \widehat L_k(G)
\, \widehat L(R) \;,
\ee
which corresponds to the k-th term in the partition functions 
(\ref{amkkahler}) on $K$.


\section{Factorization}

Having computed all the four amplitudes contributing to the anomaly,
we are now in the position of facing the interpretation in terms of 
quantum anomalies and classical inflows, and understand the mechanism 
allowing their cancellation. We will also extract all the anomalous
couplings to twisted RR fields by factorization.

\subsection{Quantum anomalies}

The anomaly arising from open string states is given by the $A$ and 
$M$ partition functions: ${\cal A}_{open} = {\cal A}_A + 
{\cal A}_M$. In total, one has:
\be
{\cal A}_{open} = \frac i{4N} \sum_{k=1}^{N-1} C_k\,
A_k(G)\Big[{\rm ch}^2_{k}(F) - {\rm ch}_{2k}(2F) \Big]
\,\widehat A(R) \,.
\label{Iopen}
\ee
The anomaly from closed string states comes instead from the $K$
and $T$ partition functions: 
${\cal A}_{closed} = {\cal A}_K + {\cal A}_T$, where ${\cal A}_T$
denotes all the contributions in (\ref{Ttot}). 
It turns out that the ${\cal A}_K$ precisely cancels 
against the untwisted RR part of ${\cal A}_T$. This reflects the fact 
that all the descendants of the anti-self-dual 4-form of the original 
Type IIB theory are projected out by the $\Omega$-projection. One is 
then left with: 
\bea
{\cal A}_{closed} &=& \frac i{2N} \sum_{k=1}^{N-1} C_k \,
\Big[\widehat A_k(G) \, \widehat G(R) 
+ \widehat G_k(G) \, \widehat A(R) \Big] \nn \\
&\;& + \frac 1{2N} \sum_{k=0}^{N-1} \sum_{l=1}^{N-1}
N_{k,l} \, {\rm ch}\,(G) \,\widehat A(R)\;.
\label{Iclosed}
\eea

The quantum anomalies (\ref{Iopen}) and (\ref{Iclosed}) can be qualitatively 
understood in their alternative interpretation as anomalies involving 
internal reparametrizations. Indeed, in that context it is easy to discuss 
the representation of each state under all the symmetries. In particular,
all the open string states and the untwisted closed string states transforms
under tangent and internal reparametrizations in a way which is dictated
essentially by dimensional reduction. This is easily made precise after 
recalling that the characteristic classes (\ref{Ahat}), (\ref{Ghat}) and 
(\ref{Lhat}) signal spinor, gravitino and self-dual representations 
under tangent reparametrizations, and similarly (\ref{Ak}), (\ref{Gk}) and 
(\ref{Lk}) correspond to spinor, gravitino and self-dual representations 
under internal reparametrizations. The open string contribution 
(\ref{Iopen}) comes clearly from a chiral spinor in $D=10$, which once 
dimensionally reduced to $D=4$ gives rise to a multiplet of chiral spinors 
transforming as an internal spinor. Similarly, the untwisted part 
(first two terms) of the closed string contribution (\ref{Iclosed}) 
come from a chiral gravitino in $D=10$, which when dimensionally 
reduced to $D=4$ gives rise to a multiplet of chiral gravitinos transforming 
as an internal spinor (first term), plus a multiplet of chiral spinors 
transforming as an internal gravitino (second term). Even the
canceled contribution of the states projected out by the orientifold
projection in summing the $K$ and $T$ surfaces can be understood. 
They come, as anticipated, from a self-dual form in $D=10$, which is 
eventually projected out, but would give rise in $D=4$ to a multiplet
of self-dual forms transforming as an internal self-dual from.
The only contribution which cannot be understood in this way is the
twisted part (third term) of (\ref{Iclosed}). It is clear that the 
correponding states must be chiral spinors, and one can argue intuitively 
that they should transform in a simpler way than untwisted fields under 
internal reparametrizations (not as tensors), since they arise at given 
fixed-points in the internal space. Indeed, it is clear from the Chern 
character in their contribution that they transform with a 
common $U(1)$ charge.

The interpretation and analysis of (\ref{Iopen}) and (\ref{Iclosed})
as sigma-model anomalies is postponed to Section 7.

\newpage

\subsection{Classical inflows}

The GS inflow, which cancels the anomalies computed in previous 
section, is given by the $t \rightarrow 0$ limit of the $A$, $M$ and $K$ 
partition functions (\ref{amkkahler}). By factorization, 
it is then possible to obtain the anomalous couplings responsible for the 
inflows. 

As in the case without composite background \cite{ss2,ss3}, the $A$, $M$ and 
$K$ partition functions have to factorize exactly. This is made possible by 
the following non-trivial identities among the characteristic classes defined 
in Section 4:
\bea
&& \sqrt{\widehat A(R)} \, \sqrt{\widehat L(R/4)} = \widehat A(R/2) \;, \\
&& \sqrt{\widehat A_{2k} (G)} \, \sqrt{\widehat L_k(G/4)} 
= \widehat A_k(G/2) \;.
\eea
Indeed, by performing suitable rescalings (allowed 
by the fact that only the 6-form component of all the polynomials is relevant)
and summing the $k$-th and the $N-k$-th terms in the sums since they 
correspond to the same closed string twisted sector, the partition functions 
(\ref{amkkahler}) can be rewritten in the factorized form
\bea
&& Z_{A} = \frac i2 \sum_{k=1}^{(N-1)/2} 
N_k \, Y_{(k)} \wedge Y_{(k)} \;, \nn \\ 
&& Z_{M} = i \sum_{k=1}^{(N-1)/2} 
N_k \, Y_{(2k)} \wedge Z_{(2k)} \;, \nn \\
&& Z_{K} = \frac i2 \sum_{k=1}^{(N-1)/2} 
N_k \, Z_{(2k)} \wedge Z_{(2k)} \;,
\label{II}
\eea
where $N_k = C_k^2$ is the number of fixed-points and
\bea
&& Y_{(k)} = \frac {\epsilon_k}{\sqrt{N}}\,
\sqrt{\left|\frac 1{C_k}\right|}\, {\rm ch}_k(\epsilon_k F)\,
\sqrt{\widehat A_k(\epsilon_k G)} \,
\sqrt{\widehat{A}(R)} \;, \raisebox{21pt}{} \nn \\
&& Z_{(2k)} = -\frac {4 \, \epsilon_{k}}{\sqrt{N}}\,
\sqrt{\left|\frac {C_{2k}}{C^2_k}\right|}\,
\sqrt{\widehat L_k(\epsilon_{2k} G/4)} \,
\sqrt{\widehat{L}(R/4)} \;. \raisebox{20pt}{} \label{anomaC}
\eea
This implies the following anomalous couplings \cite{ss3}:
\bea
&& S_{D} = \sqrt{2 \pi} \sum_{k=1}^{(N-1)/2} \sum_{i_k=1}^{N_k} 
\int C^{i_k}_{(k)} \wedge Y_{(k)} \;, \label{SD}\\
&& S_{F} = \sqrt{2 \pi} \sum_{k=1}^{(N-1)/2} \sum_{i_k=1}^{N_k} 
\int C^{i_k}_{(2k)} \wedge Z_{(2k)} \label{SF}\;.
\eea
In these couplings, $C^{i_k}_{(k)}$ denotes the sum of all the RR 
forms in the $k$-twisted sector and at the fixed-point $i_k$; it 
contains a 4-form plus a 2-form $\tilde\chi^{i_k}_{(k)}$ and its dual 
0-form $\chi^{i_k}_{(k)}$. The relevant components of the charges 
(\ref{anomaC}) are therefore the 0, 2 and 4-forms. 

\noindent
Thanks to the tadpole condition (\ref{tadpole}), all 
irreducible terms in the inflow (\ref{II}) vanish, and no unphysical 
negative RR forms propagate in the transverse channel.

The total GS couplings can be obtained by summing the D-brane 
and fixed point contributions (\ref{SD}) and (\ref{SF}), 
after sending $k$ into $2k$ in (\ref{SD}).
This is allowed for $N$ odd, and also in agreement with the fact that 
in the transverse channel one finds $k$-twisted states on $A$, 
and $2k$ twisted states on $M$ and $K$ for 
the $k$-th term in the partition function; in order to add the two 
consistently one is therefore led to the above substitution. 
Defining the quantities $X_{(2k)} = Y_{(2k)} + Z_{(2k)}$, one has
\be
S_{GS} = \sqrt{2 \pi} \sum_{k=1}^{(N-1)/2} \sum_{i_k=1}^{N_k} 
\int C^{i_k}_{(2k)} \wedge X_{(2k)} \;.
\ee
Using the explicit form (\ref{anomaC}) of the charges and the 
tadpole cancellation condition (\ref{tadpole}), one can check that 
the total charges $X^{(0)}_{(2k)}$ with respect to the RR 4-forms 
are zero, and the following results for the total charges 
$X^{(2)}_{(2k)}$ and $X^{(4)}_{(2k)}$ with respect to the RR 2-forms 
$\tilde\chi^{i_k}_{(2k)}$ and the RR 0-forms $\chi^{i_k}_{(2k)}$ are found:
\bea
X^{(2)}_{(2k)} &=& \frac {N_k^{-1/4}}{\sqrt{N}(2\pi)} \,
\Bigg\{i\,{\rm tr}(\gamma_{2k} F) + \frac 12 {\rm tr}(\gamma_{2k}) \,
\sum_{i=1}^3 \tan (\pi k v_i) \, G_i \Bigg\} \;, \label{X2} \\
X^{(4)}_{(2k)} &=& - \frac {\epsilon_{2k}\,N_k^{-1/4}}{2\sqrt{N}(2\pi)^2} \,
\Bigg\{{\rm tr}(\gamma_{2k} F^2) - \, \frac 1{32}\, {\rm tr}(\gamma_{2k}) 
\,{\rm tr}\,R^2 + i\, {\rm tr}(\gamma_{2k} F) \, 
\sum_{i=1}^3 \cot (2 \pi k v_i) \, G_i \nn \\
&\;& \hspace{75pt} - \frac 14 \, {\rm tr}(\gamma_{2k}) \, 
\Bigg[\sum_{i=1}^3 \tan^2 (\pi k v_i)\, (G_i)^2 \label{X4} \\ 
&\;& \hspace{135pt} + \,2 \sum_{i \neq j=1}^3 
\frac {\cos(2\pi k v_i)\cos(2\pi k v_j) - 1}
{\sin(2\pi k v_i)\sin(2\pi k v_j)} 
\, G_i G_j \Bigg] \Bigg\} \nn \;.
\eea
Finally, one arrives at a very simple factorized expression for the 
6-form encoding the complete sigma-gauge-gravitational anomaly 
and its opposite inflow:
\be
{\cal A}^{(6)} = {\cal I}^{(6)} = i \sum_{k=1}^{(N-1)/2} \! N_k \,
X^{(2)}_{(2k)} \wedge X^{(4)}_{(2k)} \;.
\ee


\section{Field theory outlook}

In this section, we shall address the interpretation of the results
found through the string computation within the low-energy 
supergravity. The 2-form couplings (\ref{X2}) will be responsible
for a modification of the kinetic terms of the twisted RR axions,
and will force the latter to transform non-homogeneously under gauge
and modular transformations. The 4-form couplings will then become
anomalous and generate the GS inflow required to cancel all the 
anomalies. 

In the following, we focus on $FFG_i$ and $RRG_i$ anomalies, 
since these can be compared to field theory expectations.

\vskip 9pt
\noindent 
{\bf $FFG_i$ anomalies}
\vskip 3pt 
\noindent
This kind of anomalies arise only from open string states.
To get an explicit expression from the expansion of (\ref{Iopen}), 
it is convenient to transform $k$ into $2k$ in the annulus contribution.
Using (\ref{tadpole}), one finds for the non-Abelian part:
\be
{\cal A}^{FFG_i} = -\frac {i}{2N(2\pi)^3} 
\sum_{k=1}^{N-1} C_k \, \tan (\pi k v_i)
\, {\rm tr}(\gamma_{2k} F^2) \, G_i \;.
\label{phk}
\ee
As we have seen in Section 2, $FFG_i$ anomalies are encoded in some
coefficients $b_a^i$ defined through
\be
{\cal A}^{FFG_i} = \frac i{2(2\pi)^3} b_a^i \, 
{\rm tr}(F_a^2)\, G_i \;,
\ee
where the index $a$ label the various factors of the gauge 
group. The coefficients $b_a^i$ are found to be in agreement with those
computed in \cite{iru2} for any $a,i$ and for both the $\Z_3$ and 
$\Z_7$ models. This confirms the conjectured anomaly cancellation
of mixed $FFG_i$ anomalies through a GS mechanism involving RR axions,
as proposed in \cite{iru2}. Indeed, the same anomaly polynomial
is reproduced and by factorization the expected couplings are obtained, 
{\it i.e.} the second term in (\ref{X2}). The one-forms 
$(X_{(2k)}^{(2)})^{(0)}$ modify the kinetic terms for the
axions $\chi_{(2k)}^{i_k}$. Strictly speaking it is only the combination
\be
\chi_{2k}=\frac 1{\sqrt{N_k}} \sum_{i_k=1}^{N_k} \chi_{(2k)}^{i_k}
\ee
that gets modified, since all the axions enter in a completely symmetric
way in the GS mechanism \cite{ss3}. Whereas the first term in (\ref{X2})
induces a non-homogeneous $U(1)$ transformation for $\chi_{2k}$ that eventually
leads to a Higgs mechanism through which $\chi_{2k}$ itself is eaten 
by the $U(1)$ field, the second term in (\ref{X2}) leads to a non-homogeneous 
modular transformation for $\chi_{2k}$. Note that the WZ descent for $G_i$ 
is $G_i^{(1)} = -i/2 [ \lambda^i(t^i)-\bar \lambda^i(\bar t^i) ]$,
with $\lambda^i(t^i)$ the lowest component of (\ref{lambda}).
Correspondingly, the (normalized) kinetic term for $\chi_{2k}$ will be 
invariant under sigma-model transformations if the associated
superfields $M_{2k}$ transforms, under $SL(2,R)^i$, in the following 
non-homogeneous way:
\be
M_{2k} \rightarrow M_{2k} - \frac 1{8\pi^2} \,\alpha_{2k}^i \,\lambda^i(T^i)
\;, \label{nonom}
\ee
with
\be
\alpha_{2k}^i = \frac{(2\pi )^{3/2}}{\sqrt{N}} \, N_k^{1/4} \,
{\rm tr}(\gamma_{2k}) \,\tan (\pi k v_i) \;.
\ee

\vskip 9pt
\noindent 
{\bf $RRG_i$ anomalies}
\vskip 3pt
\noindent
This kind of anomalies gets contribution both from open and closed 
string states.
Performing the same manipulation as before in the sum over $k$ 
for the annulus contribution, and summing the contributions (\ref{Iopen}) 
and (\ref{Iclosed}) from open and closed strings, one finds 
\bea
{\cal A}^{RRG_i} &=& \frac {i}{96N(2\pi)^3} \Bigg\{\sum_{k=1}^{N-1} 
C_k \, \Big[\tan (\pi k v_i) - \frac 12 \cot (\pi k v_i)\Big]
\,{\rm tr}(\gamma_{2k}) \nn \\ 
&\;& \hspace{58pt} + \sum_{k=1}^{N-1} C_k \,\cot (\pi k v_i)\, 
\Big[21 + 1 - 2 \Big(4\sin^2 (\pi k v_i) 
+ \sum_{j=1}^3 \cos (2\pi k v_j)\Big) \Big] \nn \\ 
&\;& \hspace{58pt} + \sum_{k=0}^{N-1} \sum_{l=1}^{N-1}
\, N_{k,l} \, \Bigg\} {\rm tr}\,R^2\,G_i \;. \label{RRK}
\eea
The first line comes from the open strings, and the second and 
third line from untwisted and twisted closed strings. As expected 
the untwisted RNS contribution in the first line encodes the anomaly 
of the gravitino ($21$), the dilatino ($1$), and the fermionic 
partners of the three untwisted moduli 
($-2(4\sin^2 (\pi k v_i) + \sum_j \cos (2\pi k v_j))$).
The RNS twisted sector contribution in the second line corresponds
instead to the anomaly of the neutralini. By explicit evaluation one finds
finally:
\be
{\cal A}^{RRG_i} = - \frac {i}{48(2\pi )^3}\,
{{- \, 10 \, + \, 21 \, + \, 1 \, - \, 3 \, - \, 27} 
\brack {- \, 6 \;\; + \, 21 \, + \, 1 \, - \, 1 \, - \, 21}}\, 
{\rm tr} R^2 G_i \;.
\label{RRKexp}
\ee
The coefficient in the square brackets has to be compared with 
$b^i_{grav.}=b^i_{open}+b^i_{closed}$ of \cite{lln}, the upper and
lower raws corresponding to the $\Z_3$ and $\Z_7$ models respectively.
In the notation of \cite{iru2}, the explicit form of these
coefficients is:
\bea
&& b^i_{closed} = 21 \, + \,1 + \, \delta^i_T \, + \,
\sum_\alpha \,(1+ 2\,n_i^\alpha) \;, \nn \\
&& b^i_{open} = - \, {\rm dim}\, G \,+\, \sum_{a=1}^3 \,(1+2n_i^a) \,\eta_a \;.
\label{bcoeff}
\eea
In $b^i_{closed}$, $\delta^i_T$ is the total contribution of the untwisted
moduli (nine for the $\Z_3$ model and three for the $\Z_7$ model)
and $\alpha$ runs over all the twisted massless states. These are 
assumed to have modular weight $n_i^\alpha$ as defined in Section 2,
and in \cite{lln} it was assumed that $n_i^\alpha=0$.
In $b^i_{open}$, the first term is the contribution of
the gaugini, where $G$ is the total gauge group of the model, 
$n^i_a=-\delta^i_a$ are the modular weights of the charged fields 
$C^a$ and $\eta_a$ simply counts the number of charged states belonging 
to the group $a$. Comparing the string result (\ref{RRKexp}) 
with the field theory expectations given by (\ref{bcoeff}), one finds
agreement for $b^i_{open}$ (first number in (\ref{RRKexp})) 
and for the untwisted contribution in $b^i_{closed}$ (next three numbers), 
but opposite signs for the twisted contribution (last number).
Assuming the validity of (\ref{bcoeff}), agreement 
with the string results would predict twisted modular weights 
$n_i^\alpha=-1$, $\forall i,\alpha$. 
This is in apparent contradiction with the non-homogeneous transformation
(\ref{nonom}) required for the cancellation of sigma-gauge anomalies.

\vskip 10pt

The sign that we find for the contribution of twisted modulini
is crucial for the realization of the GS anomaly cancellation mechanism,
since it directly influences the factorizability of the quantum anomaly.
We do not have a full understanding of this discrepancy; rather, we would 
like to revisit the assumptions at the origin of the above field theory 
analysis and point out a few delicate points. 
A first point to observe, in comparing string results with field 
theory expectations in $D=4$ $N=1$ models, is that 
the first are believed to be expressed in terms of linear multiplets, 
whereas the latter are often given in terms of the usual chiral multiplets,
as is the case for (\ref{bcoeff}). 
The two multiplets are related by the so called 
linear multiplet - chiral multiplet 
duality, that is basically the extension to superfields of the 
duality between a two-form and a scalar in four dimensions.
It is also known that the GS terms modify the above duality 
\cite{abd}. Correspondingly, particular attention has to be paid in 
comparing the results (\ref{RRKexp}) with field theory formulae obtained
using the chiral multiplet basis as (\ref{bcoeff}) (see for instance
footnote 7).
A second very important point is that the expression (\ref{bcoeff}) 
for the anomaly coefficients are valid only under the assumption that 
the K\"ahler potential $K^{(M)}$ for twisted fields and their modular 
transformations have the form (\ref{modw}). Unfortunately, the 
potential $K^{(M)}$ has not been computed yet in Type IIB orientifold 
models, and therefore it is not possible to verify directly these assumptions. 

We propose that the K\"ahler potential for twisted fields does in fact 
not satisfy the assumptions at the origin of (\ref{bcoeff}), so that 
the whole field-theory derivation of sigma-model anomalies, as 
reviewed in Section 2 and expressed in (\ref{bcoeff}), 
has to be revisited \cite{kiss}. 
A first possibility is that $K^{(M)}\sim (M+\bar M)^2$, as proposed 
in \cite{popp}.
This potential satisfies the assumptions behind (\ref{bcoeff}) (and leads 
to $n_i^\alpha=0$ as assumed in \cite{lln}), but only if one neglects 
the correction induced by the GS couplings (\ref{X2}) and (\ref{X4}). 
These are indeed present, as described in \cite{klein}, and it might be 
that they must be considered on equal footing with the rest of the 
potential\footnote{This seems quite strange from a string theory point 
of view, but we believe it might be 
reasonable in light of the string coupling dependence 
of the definition (\ref{Tmod}) for the $T^i$ moduli.}.
A second possibility is that $K^{(M)}$ is a different function of 
the twisted moduli, invariant under sigma-model transformations and 
the shift (\ref{nonom}), whose form does not satisfy the assumptions 
leading to (\ref{bcoeff}).

An explicit string computation of $K^{(M)}$ would therefore be
extremely interesting and could give a definite answer to the 
problems raised  above. Unfortunately, such a computation appears 
to be quite complicated.


\section{Conclusions}

In this paper, we have studied along the lines of \cite{ss2,ss3}
the pattern of sigma-gauge-gravitational anomaly cancellation in 
compact Type IIB $D=4$ $N=1$ $\Z_N$ orientifolds with $N$ odd. Our main
result is that all the anomalies are cancelled through a generalized 
GS mechanism.

The starting point of our analysis is the definition of the effective 
vertex operator corresponding to the sigma-model connection. We provided
several general arguments for identifying it with the vertex encoding 
K\"ahler deformations of the orbifold, but we were able to give only
a not completely rigorous derivation which cannot be taken as a proof.
A posteriori, this identification is strongly supported also by the results
obtained for the anomalies using this vertex. Under the assumption that
the effective vertex is indeed correct, we generalize the known results 
\cite{iru1,ss3} for gauge-gravitational anomalies and show that all 
possible sigma-gauge-gravitational anomalies are cancelled through a 
GS mechanism. This is essentially what was proposed in \cite{iru2} 
for sigma-gauge anomalies, and seems to evade the arguments of
\cite{lln} against a field theory mechanism for the cancellation
of sigma-gravitational anomalies. We interpret this discrepancy 
as evidence that the 
comparison of the string results with the field theory expectations 
is probably more subtle than expected. In particular, we propose
that the actual K\"ahler potential for twisted fields does not 
satisfy the usual assumptions made in the literature, so that 
the interpretation of our string results remains actually open.

We would like to stress that the present results imply a {\it full} 
cancellation of anomalies in all possible channels. The torus contribution 
presents a surprising cancellation and yields vanishing anomalies and 
inflows. This implies in particular that the dilaton field does not play 
any role in the GS mechanism. The annulus, M\"obius strip and Klein bottle 
contributions are instead topological, guaranteeing an exact cancellation 
between quantum anomalies and classical inflows mediated by twisted RR 
axions.


\acknowledgments

It is a pleasure to thank L.E. Ib\'a\~nez and M. Klein for continuous
support and collaboration. We would like also to thank C.P. Bachas, 
G.L. Cardoso, J.-P. Derendinger, J. Fr\"ohlich and H. Nilles for 
interesting discussions and useful comments. We acknowledge the 
Universidad Autonoma de Madrid for hospitality. C.A.S also thanks the 
Edwin Schr\"odinger Institute of Vienna and the Institute for Theoretical 
Physics of the ETH in Z\"urich. This work has been supported by the EC 
under TMR contract ERBFMRX-CT96-0045 and by the Fundamenteel Onderzoek 
der Materie (FOM).


\appendix

\section{$\vartheta$-functions}

For convenience, we introduce here a convenient notation 
for the twisted $\theta$-functions appearing in orbifold
and orientifold partition functions. In particular, in order 
to keep manifest the origin of each of these, we shall define
\bea
&& \theta_1{\alpha \brack \beta}(z|\tau) = 
\theta{\frac 12 +\alpha \brack \frac 12 + \beta}(z|\tau) \;, 
\label{theta1} \\
&& \theta_2{\alpha \brack \beta}(z|\tau) = 
\theta{\frac 12 +\alpha \brack 0 + \beta}(z|\tau) \;, 
\label{theta2} \\
&& \theta_3{\alpha \brack \beta}(z|\tau) = 
\theta{0 +\alpha \brack 0 + \beta}(z|\tau) \;, 
\label{theta3} \\
&& \theta_4{\alpha \brack \beta}(z|\tau) = 
\theta{0 +\alpha \brack \frac 12 + \beta}(z|\tau) 
\;, \label{theta4}
\eea
in terms of the usual twisted $\theta$-functions
\bea
\theta{\alpha \brack \beta}(z|\tau) = \sum_n q^{\frac 12 (n - \alpha)^2}
e^{2 \pi i (z - \beta)(n - \alpha)} \;. \label{theta}
\eea
All the properties and identities relevant to the usual $\theta$-functions
(\ref{theta}) easily translate into analogous properties of 
(\ref{theta1})-(\ref{theta4}).


\section{Anomalies in Type IIB string theory}

The cancellation of gravitational anomalies in Type IIB 
supergravity theories requires non-trivial identities 
involving the anomalies of dilatinos, gravitinos and self-dual forms. 
These are given by
\be
I_{1/2} = \widehat{A}(R) \;,\;\; I_{3/2} = \widehat{G}(R) \;,\;\;
I_{A} = - \frac 18 \widehat{L}(R) \;,\;\;
\ee
in terms of the characteristic classes (\ref{Ahat})-(\ref{Lhat}).
From a Type IIB string theory point of view, anomaly freedom is 
more manifest since the corresponding torus amplitude is perfectly 
finite. However, it is clear that in the low-energy field theory 
limit one has to reproduce in string theory the same non-trivial 
identity. This can be regarded as a technique to compute
anomalous Feynman diagrams using a string regularization.
Due to the relevance of the torus amplitude in the mixed
sigma-gauge-gravitational anomalies considered in this paper,
we find useful to report here some details on how to reproduce
in Type IIB string theory the aforementioned identity. 

As explained in Section 3 and 4, the only potentially anomalous 
contributions on the torus come from the three odd-even and the 
three even-odd spin-structures, and the total anomaly is given 
by the expression (\ref{anotoro}), in terms of the partition 
functions (\ref{part}) defined through the deformation vertices
(\ref{gravtor}) and (\ref{kator}). It turns out that the two 
partition functions (\ref{part}) will always be modular invariant, 
so that the ${\partial \cal F}_0$ component of the boundary gives 
a vanishing contribution. Moreover, on the other component 
${\partial \cal F}_\infty$ of the boundary, the odd-even and even-odd
partition functions become equal and  sum. In the following, 
we will therefore restrict to the odd-even spin-structures.

\vskip 9pt
\noindent 
{\bf $D=10$}
\vskip 3pt 
\noindent
In the ten dimensional case, the partition functions (\ref{part}) are 
particularly easy to compute. 
One gets
\be
Z_T^{S_\alpha} = \frac 14 \prod_{a=1}^5 \left[ \frac{ix_a}
{\theta_1(ix_a/\pi|\tau)}\,\theta_\alpha(ix_a/\pi|\tau) \right] 
\frac{\eta^3(\tau )}{\theta_\alpha(0|\tau)}\;.
\ee
Here $\alpha=2,3,4$ represent respectively the RR, ${\rm RNS}_+$ and
${\rm RNS}_-$ spin-structures,
the factor of $1/4$ is due to the left and right GSO projections
and $x_a=\lambda_a/2\pi$, in terms of the skew eigenvalues $\lambda_a$
of the gravitational curvature $R$. 
The first fraction is the contribution 
of the bosonic and fermionic fields, whereas the last fraction is due to 
ghosts and superghosts. Taking the limit $\tau_2\rightarrow \infty$, one 
obtains in the RR spin-structure
\be
Z_T^{RR} \rightarrow \frac 18 \prod_{a=1}^5 \frac{x_a}{\tanh x_a} \;.
\ee
In the ${\rm RNS}_\pm$ spin-structures, similarly
\be
Z_T^{RNS} =  Z_T^{RNS_+} - Z_T^{RNS_-} \rightarrow 
\prod_{a=1}^5 \frac{x_a/2}{\sinh x_a/2}
\Big(2\sum_{b=1}^5 \cosh x_b -2\Big) \;,
\label{NS}
\ee
where we rescaled by a factor of 2 the $x_a$'s, 
exploiting the fact that only the 12-form of (\ref{NS}) is relevant.
Notice that the leading ``tachyonic'' terms in $Z_T^{RNS_\pm}$ cancel 
in the combination $Z_T^{RNS_+} - Z_T^{RNS_-}$. 
By summing the three contributions one finds as expected the anomaly 
of an anti-chiral gravitino and of a chiral dilatino from the 
RNS/NSR sector and that of an (anti)self-dual tensor from the RR sector. 
In total, one gets
\be
I_{T} =  - I_{3/2} + I_{1/2} - I_{A} = 0 \;,
\ee
ensuring the absence of pure gravitational anomalies in $D=10$ 
Type IIB supergravity and superstring theory \cite{agw}.

\vskip 9pt
\noindent
{\bf $D=6$ on $T^4/\Z_N$}
\vskip 3pt
\noindent
As usual in orbifold theories, the partition functions (\ref{part})
contain a sum over orbifold twisted sectors $l$, as well
as a projection on $\Z_N$-invariant states; see (\ref{T234}). 
In the following, we will further distinguish between the 
contributions coming from untwisted and twisted sectors. The twist
vector is $v_i=(1/N,-1/N)$, $C_k = \prod_i (2\sin (\pi k v_i))$,
and $N_{k,l}$ are the number of points that are at the same 
time $k$ and $l$-fixed. The total partition function is 
\be
Z_T= \sum_\alpha (-)^\alpha \sum_{l=0}^{N-1} Z_T^{S_\alpha \, (l)} \;,
\ee
where  
\be
Z_T^{S_\alpha \,(l)} = \frac 1{4N} \sum_{k=0}^{N-1} N_{k,l} 
\prod_{i=1}^{2} \frac{\theta_\alpha{lv_i \brack kv_i}(0|\tau)}
{\theta_1{lv_i \brack kv_i}(0|\tau)}
\prod_{a=1}^3 \left[\frac{ix_a}{\theta_1(ix_a/\pi|\tau)} 
\,\theta_\alpha(ix_a/\pi|\tau) \right] 
\frac{\eta^3(\tau)}{\theta_\alpha(0|\tau)} \;.
\ee
In the $\tau_2 \rightarrow \infty$ limit, one finds in the RR 
spin-structures:
\bea
&& Z_T^{RR(0)} \rightarrow \frac 1{8N} \sum_{k=0}^{N-1} C_{2k} \, 
\prod_{a=1}^3 \frac{x_a}{\tanh x_a} \;, \nn \\
&& Z_T^{RR(l\neq 0)} \rightarrow \frac 1{8N}\sum_{k=0}^{N-1} N_{k,l} 
\prod_{a=1}^3 \frac{x_a}{\tanh x_a} \;.
\eea
One can easily check that for any $N=2,3,4,6$, the total is given by
\be
Z_T^{RR} \rightarrow 2\,\prod_{a=1}^3 \frac{x_a}{\tanh x_a} \;.
\label{ZR}
\ee
In the ${\rm RNS}_{\pm}$ spin-structures one has
to pay particular attention in taking the limit,
because when $l = N/2$, the fields in the internal directions have zero 
modes. One finds the following results:
\bea
&& Z_T^{RNS_\pm(0)} \rightarrow \pm \frac 1{2N} \sum_{k=0}^{N-1}
C_k \prod_{a=1}^3 \frac{x_a/2}{\sinh x_a/2} 
\Big(2 \sum_{b=1}^3 \cosh x_b -2 + \sum_{i=1}^{2} (2 \cos 2\pi k v_i)\Big) 
\;, \hspace{-10pt} \nn \\
&& Z_T^{RNS_\pm(l \neq 0,N/2)} \rightarrow 
\pm \frac 1{N} \sum_{k=0}^{N-1} N_{k,l}
\prod_{a=1}^3 \frac{x_a/2}{\sinh x_a/2} \;, \nn \\
&& Z_T^{RNS_+(N/2)} \rightarrow \frac 1{2N} \sum_{k=0}^{N-1} N_{k,N/2}
\prod_{a=1}^3 \frac{x_a/2}{\sinh x_a/2}
\prod_{i=1,2}(2\cos \pi k v_i)^2 \nn \\
&& Z_T^{RNS_-(N/2)} \rightarrow -\frac 1{2N} \sum_{k=0}^{N-1} N_{k,N/2}
\prod_{a=1}^3 \frac{x_a/2}{\sinh x_a/2} 
\prod_{i=1,2}(2\sin \pi k v_i)^2 \;.
\eea
We omitted the leading ``tachyonic'' term that, as in the previous case, will
cancel in taking the sum $Z_T^{RNS_+} - Z_T^{RNS_-}$.
One can easily verify that the total result in the RNS sectors, obtained
by summing over the two ${\rm RNS}_\pm$ contributions and over all twisted 
and untwisted sectors, is the same for any $N=2,3,4,6$ and given by
\be
Z_T^{RNS} = - \, 2\, \prod_{a=1}^3 \frac{x_a/2}{\sinh x_a/2} 
\Big(2 \sum_{b=1}^3 \cosh x_b - 22\Big) 
\label{ZNS}
\ee
Putting all together, one gets finally
\be
I_{T} =  2 \, ( I_{3/2} - 21 I_{1/2} - 8 I_A)  = 0 
\ee
ensuring the absence of purely gravitational anomalies 
in Type IIB theory on $T^4/\Z_N$.


\section{Vanishing of the torus amplitude}

We show here that the whole 6-form component of the torus amplitude,
including $G^3$ anomalies, vanishes.
For the $RRG_i$ terms, one gets 
\bea
Z^{R^2 G}_T &=& \frac {i}{96N(2\pi)^3} \sum_{i=1}^3 \left\{
-\,4\, \sum_{k=1}^{N-1} C_{2k} \sin^{-1} (2\pi k v_i) \right. 
\nn \\ &\;& \hspace{74pt} + \sum_{k=1}^{N-1} C_k \cot (\pi k v_i) 
\Big[21 + 1 - 2 \Big(4 \sin^2 (\pi k v_i) +
\sum_{j=1}^3 \cos (2\pi k v_j)\Big) \Big] \nn \\ &\;& \hspace{73pt} 
\left. + \sum_{k=0}^{N-1} \sum_{l=1}^{N-1} N_{k,l} \right\} 
{\rm tr}\,R^2\,G_i \nn \\ &=& -\frac {i}{48(2\pi)^3} \, 
{{8 \, + \, 21 \, + \, 1 \, - \, 3 \, - \, 27} 
\brack {0 \, + \, 21 \, + \, 1 \, - \, 1 \, - \, 21}} \, 
{\rm tr}\,R^2\, \Big(\sum_{i=1}^3 G_i\Big) \nn \\
&=& 0 \;,
\eea
where we reported in square bracket the explicit values for both
the $\Z_3$ (up) and $\Z_7$ (down) orientifolds.

Consider next the $G_i G_j G_p$ terms. 
The RR twisted contributions vanish as before, whereas the
RNS twisted ones are present and can be easily read from the last line
of (\ref{Ttot}). On the contrary, the untwisted RR and RNS contributions
requires more work. However, one can now put to zero the gravitational 
curvature. By doing so, the contribution of the superghosts cancels 
that of one of the two complex spacetime fermions in (\ref{T})
and one can therefore use the Riemann identity to simplify the result.
In the $\tau_2\rightarrow\infty$ limit, one has then:
\bea
&& \frac{8 \prod_i \cos (\pi k v_i + G_i/2\pi) 
- 2 \sum_i \cos 2(\pi k v_i + G_i/2\pi) - 2}
{8 \prod_i \sin (\pi k v_i + G_i/2\pi)} \nn \\
&& \hspace{70pt} = -2 \sin \Big(\sum_{p=1}^3 G_p/4\pi\Big) 
\prod_{i=1}^3 \frac {\sin \Big[(\pi k v_i + G_i/2\pi) 
- (\sum_p G_p/4\pi)\Big]} {\sin (\pi k v_i + G_i/2\pi)} \nn \;.
\eea
The relevant cubic term of the partition function are now easily computed,
and one finds:
\bea
Z^{G^3}_T &=& \frac i{24N(2\pi)^3} \Bigg\{\sum_{k=1}^{N-1} N_k \,
\Bigg[\Big(3\sum_{i=1}^3 \prod_{j \neq i = 1}^3 \cot (\pi k v_i) - 5 \Big)
\Big(\sum_{p=1}^3 G_p\Big)^3 \nn \\ &\;& \hspace{105pt}
+ \, 6 \sum_{i=1}^3 \sin^{-2}(\pi k v_i) \, G_i 
\Big(\sum_{p=1}^3 G_p\Big)^2 \Bigg]
\nn \\ &\;& \hspace{60pt} -\,2 \sum_{k=0}^{N-1} 
\sum_{l=1}^{N-1} N_{k,l} \, \Big(\sum_{p=1}^3 G_p\Big)^3 \Bigg\} 
\nn \\ &=&  \frac i{4N(2\pi)^3} \, 
{9 \, - \, 15 \, + \, 24 \, - \, 18 \brack 
3 \, - \, 5 \;\, \, + \, 16 \, - \, 14} \, 
\Big(\sum_{p=1}^3 G_p\Big)^3 \nn \\ &=& 0 \;.
\eea


\end{document}